\documentclass[final,1p,times,twocolumn]{elsarticle}
\usepackage{epsfig}
\usepackage{amssymb}
\usepackage{amsmath}
\usepackage{subcaption}
\usepackage{longtable}
\journal{Nuclear Physics B}
\usepackage{xcolor}

\DeclareMathAlphabet{\oldcal}{OMS}{cmsy}{m}{n}
\newcommand{\bigo}[1]{\oldcal{O}\left(#1\right)}
\usepackage{hyperref}
\usepackage{multirow}

\usepackage[colorinlistoftodos]{todonotes}

\begin{document}

\title{Six-loop $\varepsilon$ expansion study of three-dimensional $O(n)\times O(m)$ spin models}

\author[label1]{M.\,V.\,Kompaniets}

\author[label1]{\corref{cor1}A.\,Kudlis}
\ead{andrewkudlis@gmail.com} 
\cortext[cor1]{Corresponding author}

\author[label1]{A.\,I.\,Sokolov}

\address[label1]{Saint Petersburg State University, 7/9 Universitetskaya Embankment, St. Petersburg, 199034 Russia}

\date{\today}

\begin{abstract}
The Landau-Wilson field theory with $O(n)\times O(m)$ symmetry which describes 
the critical thermodynamics of frustrated spin systems with noncollinear and noncoplanar ordering 
is analyzed in $4 - \varepsilon$ dimensions within the minimal subtraction scheme in the six-loop 
approximation. The $\varepsilon$ expansions 
for marginal dimensionalities of the order parameter 
$n^H(m,4-\varepsilon)$, $n^-(m,4-\varepsilon)$, $n^+(m,4-\varepsilon)$ separating different regimes 
of critical behavior are extended up to $\varepsilon^5$ terms. Concrete series with coefficients 
in decimals are presented for $m=\{2, \dots, 6\}$. 
The \textit{diagram of stability} of 
nontrivial fixed points, including the chiral one, in $(m,n)$ plane is constructed by means of summing up of corresponding $\varepsilon$ expansions using various resummation techniques. Numerical estimates of the chiral critical 
exponents for several couples $\{m,n\}$ are also found. Comparative analysis of our results with 
their counterparts obtained earlier within the lower-order approximations and by means of alternative 
approaches is performed. It is confirmed, in particular, that in physically interesting cases $n=2, m=2$ 
and $n=2, m=3$ phase transitions into chiral phases should be first-order. 
\end{abstract}

\begin{keyword}
renormalization group, chiral model, multi-loop calculations, marginal dimensionalities, 
$\varepsilon$ expansion, critical exponents, frustrated spin systems.

\MSC{82B28}
\end{keyword}

\maketitle

\section{Introduction}

Compared to conventional (anti)ferromagnets, where ordering occurs in a collinear manner, phase transitions in systems with frustration are much less studied. 
The source of such frustration can be the geometric features of crystal lattice as well as the nature 
of the interactions between nearest and next-nearest spins. In case the number of spin components 
is greater or equal to two $(n \geqslant 2)$ the presence of frustration leads to a noncollinear 
and a noncoplanar spin ordering. As prime examples of systems with mentioned non-trivial spin structures  
stacked triangular antiferromagnets (STA) and helical magnets (HM) may be considered. In the course 
of studying the critical properties of such systems, the main question arises: which type of 
a phase transition takes place in real materials -- whether it has to be a continuous one with 
non-standard set of critical exponents corresponding to so-called \textit{chiral} universality 
class or a phase transition should be first-order.

There are many experimental works dedicated to the frustrated spin systems where they were studied by means of optical second-harmonic spectroscopy, measurements of magnetization, susceptibility and specific heat, 
resistivity measurements, synchrotron X-rays and neutron diffraction, M\"{o}ssbauer technique etc. \cite{LOTTERMOSER20011131,Brunt2018,doi:10.1143/JPSJ.69.33,doi:10.1143/JPSJ.70.3068,PhysRevB.67.104431,PhysRevLett.83.2065,MITSUDA19991249,PhysRevB.62.8983,PhysRevB.66.184432,PhysRevB.66.052405,doi:10.1143/JPSJ.75.023602,TAKEUCHI20001527,ECKERT1976911,PhysRev.162.315,LEDERMAN19741373,PhysRevB.67.094434,doi:10.1143/JPSJ.79.011003,PhysRevB.99.094408,Ranjith_2017,PhysRevB.96.224427,PhysRevB.93.024412,LOH1974357,qubs2030016}. The results of these experiments turn out to be rather contradictory. For materials with the same symmetry the critical exponents measured are noticeably scattered, and even the order of the phase transition was found to be different for the closely related substances. 

As for theoretical studies, there are a good number of works investigating the critical properties of frustrated magnets. Here we note only some of key points in the development of the study with emphasis on a field theoretical (FT) approach; those interested in more details and the history of the problem may be addressed to comprehensive reviews \cite{Diep1994,doi:10.1139/p97-007,Kawamura_1998,PELISSETTO2002549,diep2004frustrated,PhysRevB.69.134413}.

The first attempts to analyze the critical behavior of the frustrated spin systems by means of renormalization group (RG) approach were undertaken about four decades ago \cite{PhysRevB.13.5086,Garel_1976,PhysRevB.25.1969}. These works were intended to explain the critical properties of rare-earth magnets such as Dy, Ho and Tb. It was realized that for proper description of the systems with nontrivial ordering within FT approach an account for the $O(n)$-symmetric self-interaction is not sufficient. To cope this problem the Landau-Wilson (LW) action with $O(m)\times O(n)$ symmetry was suggested which is written down in modern notations, say, in \eqref{H}. The presence of such a specific symmetry being natural for real systems was shown to lead to the critical behavior which is different from that described by the standard ($O(n)$-symmetric) universality class \cite{doi:10.1063/1.338936,doi:10.1143/JPSJ.54.3220,doi:10.1143/JPSJ.55.2095}.   

The reasonableness of studying certain classes of universality depends on whether the corresponding regimes of critical behavior are stable or not. The stability in turn is determined by the structure of the renormalization group (RG) flows. For physically interesting values of $m$ and $n$ relevant phase portraits were analyzed. In the case $m\geqslant 3$ the results given by all FT approaches -- $\varepsilon$ expansion \cite{doi:10.1143/JPSJ.59.2305,PELISSETTO2001605,CALABRESE2004568}, $1/n$ expansion \cite{PELISSETTO2001605,GRACEY2002433}, 3D RG technique \cite{PhysRevB.68.104415}, $2-\varepsilon$ machinery \cite{PhysRevLett.64.3175,AZARIA1993485}, pseudo-$\varepsilon$ expansion \cite{CALABRESE2004568}, "exact" (functional, non-perturbative) renormalization group (ERG) 
\cite{PhysRevB.61.15327} -- are in favor of existence of some upper bound (marginal) value of $n$, $n^+$, such that for $n > n^+$ the system 
demonstrates continuous phase transition with specific set of critical exponents while for $n < n^+$ the transition is first-order. 

For $m=2$ the situation is more complicated\footnote{For convenience of reader notations of physically interesting quantities were chosen in a similar way as in \cite{PELISSETTO2001605,CALABRESE2004568}.}\cite{PELISSETTO2001605,GRACEY2002433,PhysRevLett.64.3175,AZARIA1993485,ANTONENKO1995161,PhysRevB.49.15901,loison2000,PhysRevB.63.140414,PhysRevB.66.180403,PhysRevB.68.094415,PhysRevB.64.094406,Ono_1999,ONO1998735,PhysRevB.70.174439,CALABRESE2005550}. For the same values of $n$ the results obtained by different approaches are controversial \cite{PhysRevLett.84.5208,PhysRevB.67.134422,doi:10.1142/S0217751X01004827}. Some materials with identical symmetry, being expected to belong to the same universality class described by action \eqref{H} under physical values of $n$, demonstrate considerable scattering of critical exponents measured in different samples \cite{PhysRevB.67.224408,PhysRevLett.88.027203,PhysRevB.64.100402,PhysRevLett.85.3942,PhysRevB.65.094403}. An interesting picture was obtained within 3D RG analysis: there is the value $N_{c2}<n^+$ such that the systems with $n<N_{c2}$ have to demonstrate the scaling behavior \cite{PhysRevB.68.094415}. According to the six-loop 3D RG analysis the physical values $n=2, 3$ are covered by the last case \cite{PhysRevB.63.140414,PhysRevB.66.180403}. The critical exponents obtained within 3D RG approach \cite{PhysRevB.63.140414,PhysRevB.65.020403} are in a bad agreement with experimental results. 
Regarding ERG there were found no stable fixed points within this approach signaling about realization of first-order phase transition \cite{PhysRevB.61.15327,PhysRevLett.84.5208,doi:10.1142/S0217751X01004827}.
Contradictory results are observed not only within field-theoretical methods. For STA Monte Carlo calculations performed by different groups argues both in favor of continuous phase transition \cite{PhysRevB.67.184407,PhysRevB.54.4165,PhysRevB.50.6854,PhysRevB.50.16453,1910.13112} and first-order one \cite{s100510050497,s100510050113,PhysRevE.78.031119,doi:10.1063/1.2837281}. It is worth noting here that systems can undergo first-order phase transition even if some stable fixed points exist on RG flows -- this occurs when initial (bare) values of 
couplings lie outside their regions of attraction. 

The realization of this work has become particularly relevant due to the recently obtained six-loop RG expansions for $O(n)$-symmetric model \cite{KP17}. These results allow us to extend obtained fifteen years ago five-loop $\varepsilon$ expansions for $O(m)\times O(n)$-symmetric model \cite{CALABRESE2004568} to the six-loop order. It is expected that numerical estimates obtained within this approximation will be a high priority with respect to determination of accurate values of marginal dimensionalities $n^+$, $n^-$ and $n^H$.   

The paper is organized as follows. In Section 2 the model and renormalization scheme employed  
are described. In Section 3 all the quantities of interest including RG functions, fixed points and marginal dimensionalities are calculated. The numerical estimates of $n^{+}(m,3)$, $n^{-}(m,3)$ and $n^{H}(m,3)$ for $m=\{2,\dots,6\}$ are presented in Section 4. There are also the results concerning the critical exponents for physically interesting couples $\{m,n\}$. In Section 5 the numbers obtained in this work, found earlier within the lower-order approximations and by means of alternative 
approaches are compared and analyzed. In the last section a summary of the main results is presented.
\section{Model and renormalization}
\subsection{Model}
The critical behavior of chiral systems within FT approach is described by the $O(n)\times O(m)$-symmetric LW action with two coupling constants \cite{Garel_1976,doi:10.1063/1.338936,Jones_1976,Bailin_1977,doi:10.1143/JPSJ.55.2157,PhysRevB.38.4916,PhysRevB.42.2610}:
\begin{equation}
S = \int d^D{x}\Biggl\{\frac{1}{2} \left[(\partial \varphi_{0\alpha i }) ^2+ m_0^2  \varphi_{0\alpha i}^2\right] + \frac{1}{4!}\left[ g_{01}T^{(1)}_{\alpha i, \beta j, \gamma k, \delta l} +g_{02}T^{(2)}_{\alpha i,\beta j,\gamma k,\delta l} \right] \varphi_{0\alpha i} \varphi_{0\beta j}  \varphi_{0\gamma k } \varphi_{0\delta l} \Biggr\}, \label{H}
\end{equation}
where $\varphi_{0\alpha i}$ is $m$-size set of $n$-component vector fields $(\alpha\in\{1,\dots m\}$, $i\in\{1,\dots n\})$, $g_{01}$ and $g_{02}$ are bare coupling constants, $m_0$ is a bare mass being proportional to $T-T_0$, where $T_0$ is mean-field transition temperature. The tensor factors $T^{(1)}$ and $T^{(2)}$ entering the $O(mn)$-symmetric and \textit{chiral} terms respectively are as follows
\begin{gather}
T^{(1)}_{\alpha i,\beta j,\gamma k, \delta l} = \frac{1}{3}(\delta_{\alpha i, \beta j} \delta_{\gamma k,  \delta l} +\delta_{\alpha i, \gamma k} \delta_{\beta j, \delta l}+\delta_{\alpha i, \delta l} \delta_{\gamma k,\beta j}), \quad \delta_{\alpha i, \beta j} = \delta_{\alpha \beta} \delta_{ij},\nonumber \\
T^{(2)}_{\alpha i,\beta j,\gamma k, \delta l} = \frac{1}{6}\left[\delta_{\alpha \beta} \delta_{\gamma \delta}(\delta_{ik}\delta_{jl}+\delta_{il}\delta_{jk})+\delta_{\alpha \gamma} \delta_{\beta \delta}(\delta_{ij}\delta_{kl}+\delta_{il}\delta_{jk})+\delta_{\alpha \delta}\delta_{\beta \gamma}(\delta_{ij} \delta_{kl}+\delta_{ik}\delta_{jl})\right] -T^{(1)}_{\alpha i,\beta j,\gamma k, \delta l}.
\label{T12def}
\end{gather}
In particular, 
\begin{gather}
T^{(1)}_{\alpha i,\beta j,\gamma k, \delta l}T^{(1)}_{\alpha i,\beta j,\gamma k, \delta l} = \frac{mn(mn+2)}{3}, \qquad 
T^{(1)}_{\alpha i,\beta j,\gamma k, \delta l}T^{(2)}_{\alpha i,\beta j,\gamma k, \delta l}= -\frac{mn(m-1)(n-1)}{3}, \nonumber \\ T^{(2)}_{\alpha i,\beta j,\gamma k, \delta l}T^{(2)}_{\alpha i,\beta j,\gamma k, \delta l} = \frac{mn(m-1)(n-1)}{2}=-\frac{3}{2}T^{(1)}_{\alpha i,\beta j,\gamma k, \delta l}T^{(2)}_{\alpha i,\beta j,\gamma k, \delta l}.  
\label{T12exmpls}
\end{gather}

Assuming $n\geqslant m$, to provide a noncollinear ordering and positive definiteness of action \eqref{H} it is necessary to impose on the coupling constants the following conditions: $0<g_{02}/g_{01}<m/(m-1)$\cite{PhysRevB.49.15901,PELISSETTO2001605,CALABRESE2004568}. In the case of negative $g_{02}$ the equation \eqref{H} describes the magnets with simple unfrustrated ordering and with sinusoidal spin structure\cite{PhysRevB.38.4916,PhysRevB.42.2610}. For $m=2$ the model \eqref{H} corresponds to systems with non-collinear but coplanar ordering. Physically most interesting of them are XY $(n=2)$ and Heisenberg $(n=3)$ frustrated antiferromagnets. In case of $m\geqslant 3$ the action describes the critical behavior of magnets with non-coplanar ordering \cite{doi:10.1143/JPSJ.59.2305}.

\begin{figure}[h!]
\centering
\includegraphics[width=6cm]{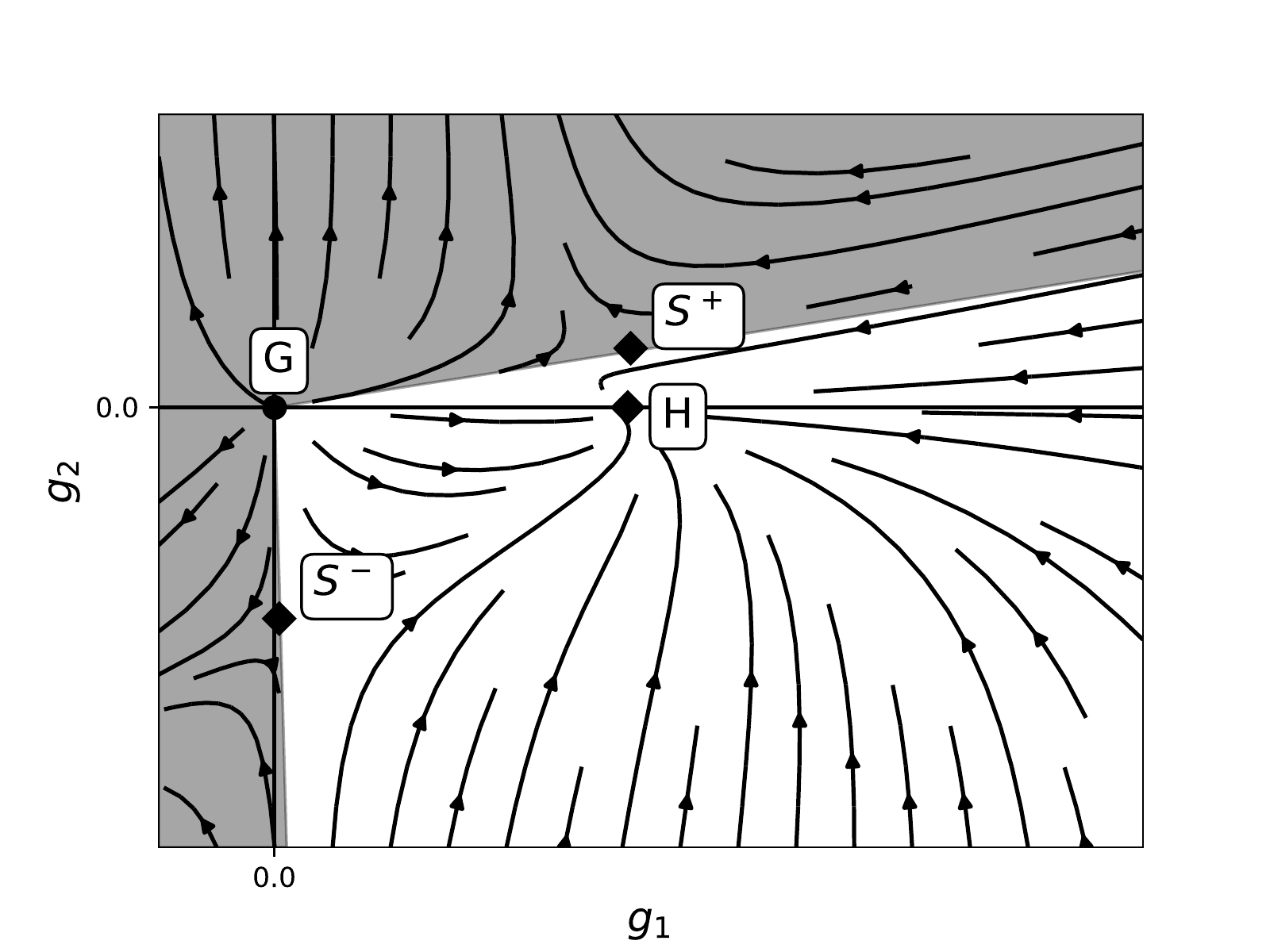}
\includegraphics[width=6cm]{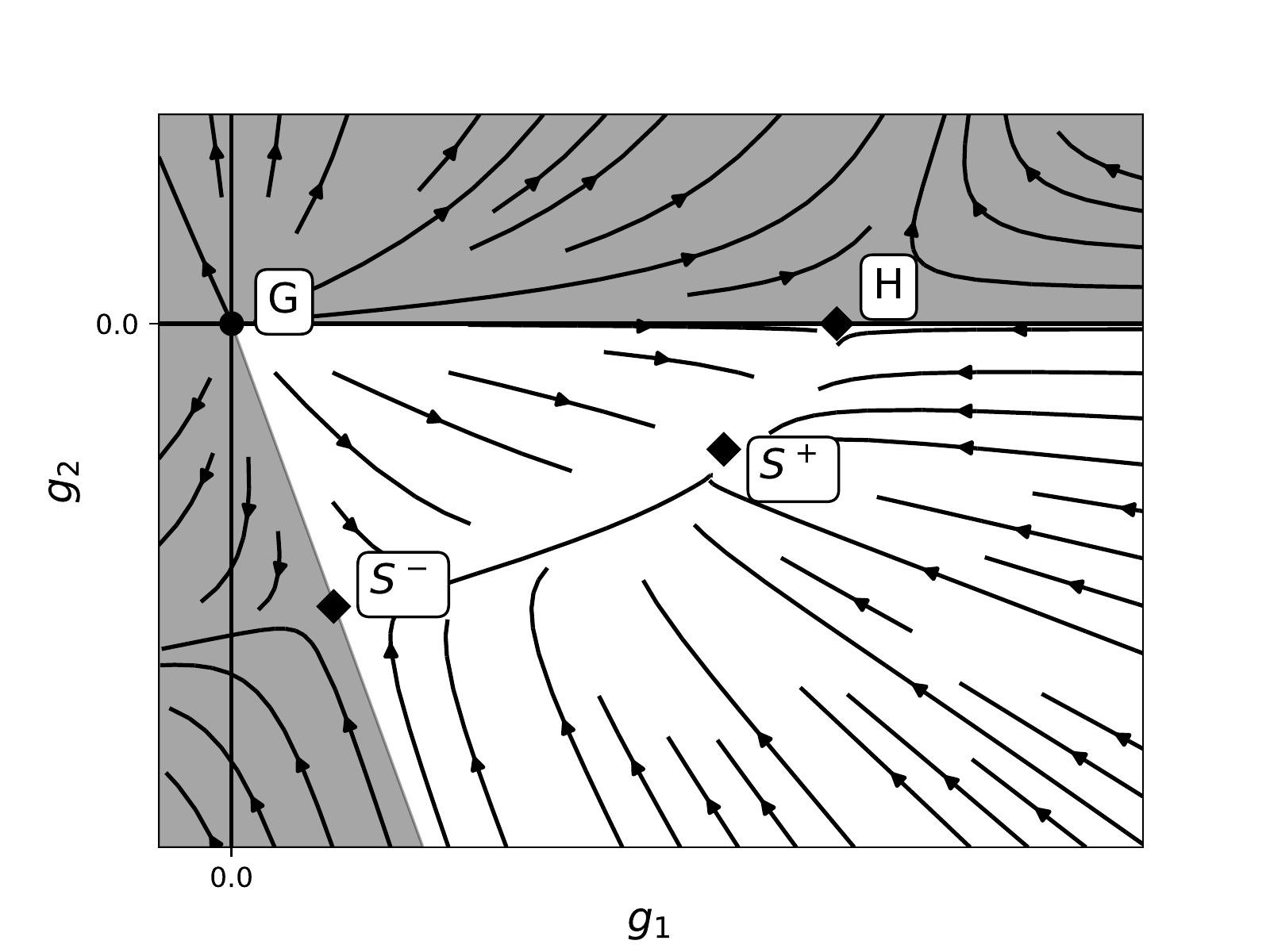}
\includegraphics[width=6cm]{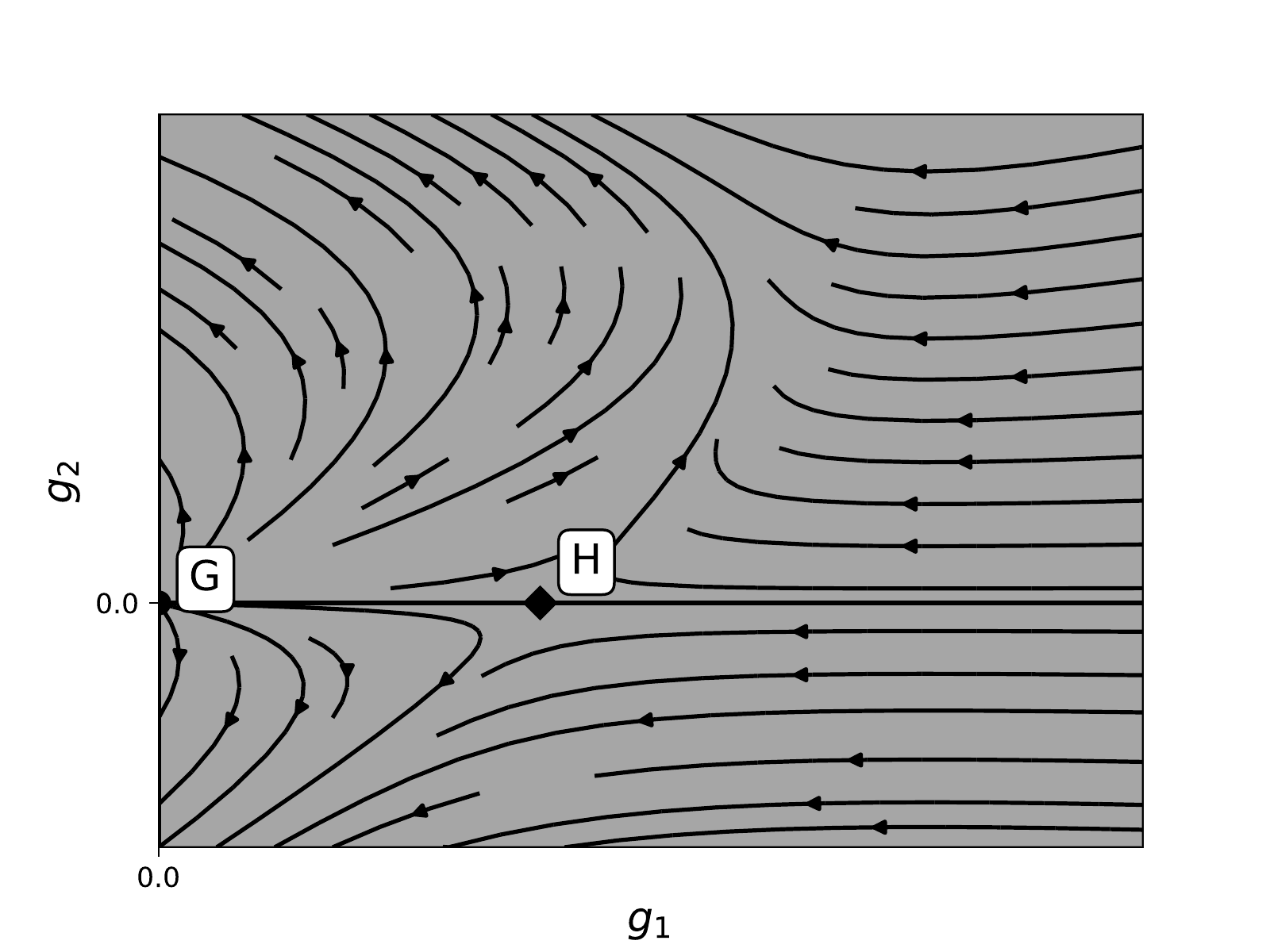}
\includegraphics[width=6cm]{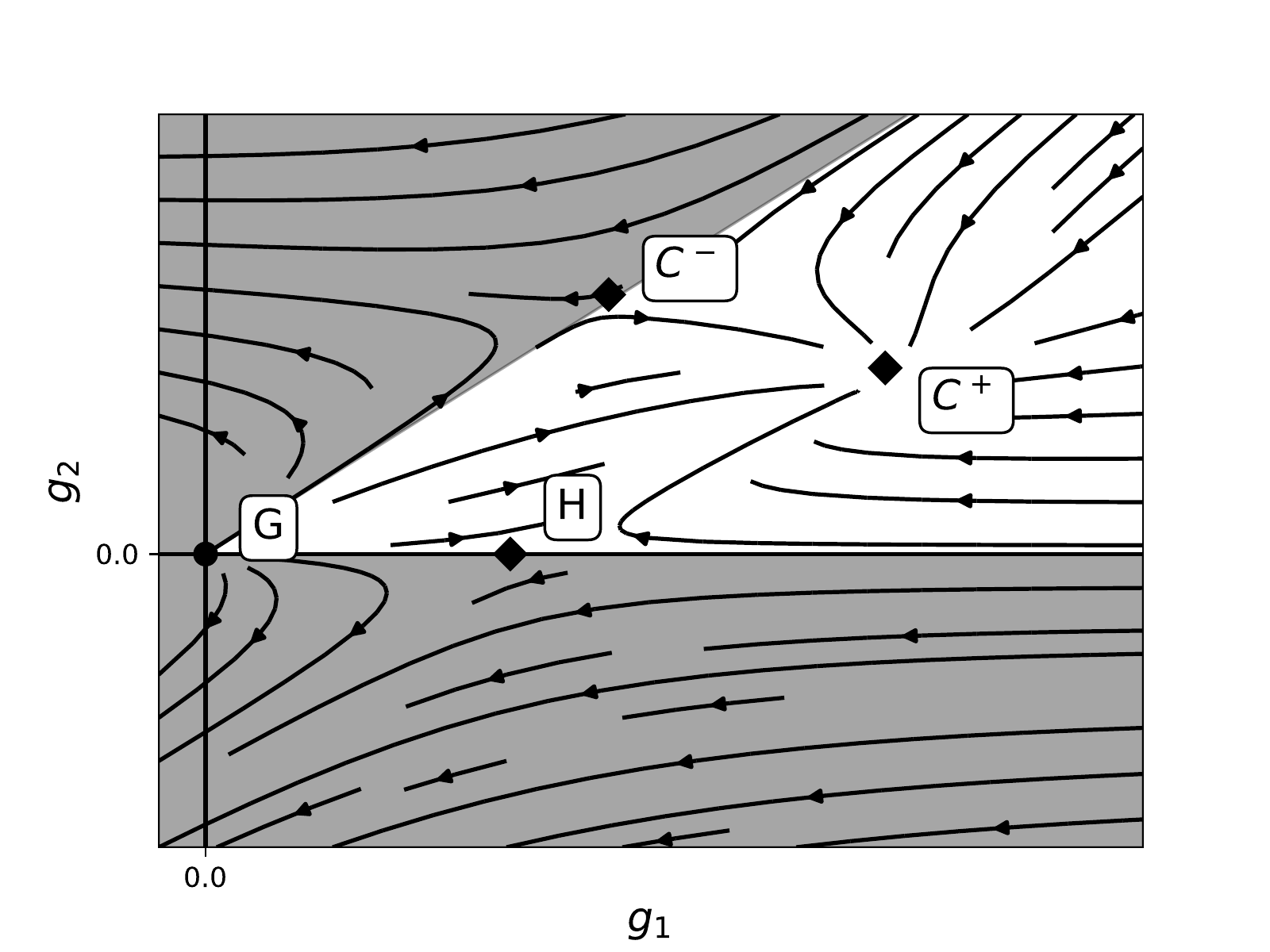}
\caption{RG flows of renormalized coupling constants. Upper left picture corresponds to $n < n^H$, upper right one -- to $n^H < n < n^-$, lower left picture corresponds to $n^- < n < n^+$ and lower right one -- to $n > n^+$. Symbols in boxes mark Gaussian ($G$), chiral ($C^+$), antichiral ($C^-$), sinusoidal ($S^+$), antisinusoidal ($S^-$) and Heisenberg ($H$) FPs. White areas correspond to regions of attraction of stable fixed points where the system demonstrates a scaling behavior.} 
\label{fig:rgflow}
\end{figure}

In the critical region, the analysis of RG flows demonstrates the presence of four different fixed points (FPs). Two of them always exist -- \textit{Heisenberg} or $O(mn)$-\textit{symmetric} $(g_{02}=0)$ and \textit{Gaussian} $(g_{01}=g_{02}=0)$ ones, while appearing of two others 
depends on the value of $n$.  For $m\geqslant1$ there are four particular cases\cite{PhysRevB.38.4916}:
\begin{enumerate}
    \item if $n<n^H(m,d)$ there are four FPs with $O(mn)$-symmetric one being stable;
    \item if $n^H(m,d)<n<n^-(m,d)$ and $m<7$ there are four FPs as well, but instead of $O(mn)$-symmetric 
    FP the anisotropic (\textit{sinusoidal}) one acquires the stability for negative $g_{02}$, while for positive values of this coupling phase transition is expected to be first-order;
    \item for $n^-(m,d)<n<n^+(m,d)$ only Gaussian and Heisenberg FPs exist and, since both are unstable, the system is expected to undergo first-order phase transition for any values of bare couplings;
    \item if $n>n^+(m,d)$ there are four fixed points with chiral one being stable and governing the chiral critical behavior.
\end{enumerate}
All the regimes described are depicted in the Figure \ref{fig:rgflow}.

\subsection{Renormalization}

The model studied is known to be multiplicatively renormalizable\footnote{The renormalization procedure is similar to that performed by the authors for another model with two coupling constants\cite{ADZHEMYAN2019332}.}. The bare quantities $g_{10},g_{20}, m_0^2, \varphi_0$ can be expressed via the renormalized ones $g_{1},g_{2}, m^2, \varphi$ by means of the following relations:
\begin{gather}
m_0^2= m^2 Z_{m^2}, \qquad  g_{01} = g_1 \mu^{\varepsilon}Z_{g_1}, \qquad g_{02} = g_2 \mu^{\varepsilon}Z_{g_2}, \qquad \varphi_0 = \varphi Z_{\varphi} ,\nonumber \\
Z_1=Z_\varphi^2 ,  \qquad      Z_2=Z_{m^2} Z_\varphi^2, \qquad Z_3=Z_{g_1} Z_\varphi^4 ,\qquad Z_4=Z_{g_2} Z_\varphi^4. 
\end{gather}
In terms of renormalized parameters the action \eqref{H} acquires the following form
\begin{gather}
S^R = \int d^D{x}\Biggl\{\frac{1}{2} \left[Z_1(\partial \varphi_{\alpha i }) ^2 + Z_2 m^2  \varphi_{\alpha i}^2\right] + \frac{1}{4!}\left[Z_3 g_{1}\mu^{\varepsilon}\;T^{(1)}_{\alpha i, \beta j, \gamma k, \delta l} + Z_4 g_{2}\mu^{\varepsilon}\;T^{(2)}_{\alpha i,\beta j,\gamma k,\delta l} \right] \varphi_{\alpha i} \varphi_{\beta j}  \varphi_{\gamma k } \varphi_{\delta l} \Biggr\}, \label{HR}
\end{gather}
where $\mu$ is an arbitrary mass scale introduced to make renormalized couplings $g_1$ and $g_2$ dimensionless. Use of RG constants eliminates all the divergences, making renormalized Green functions free of them. The multiplicative renormalizability of the model allows to limit ourselves by removal the divergences only in two- and four-point one-particle irreducible Green functions:
\begin{gather}
\Gamma^{(2)}_{\alpha i, \beta j} = \Gamma^{(2)}\delta_{\alpha i, \beta j}, \qquad \Gamma^{(4)}_{\alpha i,\beta j,\gamma k,\delta l} = \Gamma^{(4)}_1 T^{(1)}_{\alpha i,\beta j,\gamma k,\delta l}+\Gamma^{(4)}_2 T^{(2)}_{\alpha i, \beta j,\gamma k,\delta l},
\end{gather} 
\begin{gather}
\Gamma_1^{(4)} = \frac{9 \left(T^{(1)}_{\alpha i, \beta j, \gamma k, \delta l}+\frac{2}{3} T^{(2)}_{\alpha i, \beta j, \gamma k, \delta l}\right)}{m(m+2)n(n+2)} \Gamma^{(4)}_{\alpha i, \beta j, \gamma k, \delta l}, \quad
\Gamma_2^{(4)} = \frac{6\left(T^{(1)}_{\alpha i, \beta j, \gamma k, \delta l}+\frac{2+mn}{(n-1)(m-1)}T^{(2)}_{\alpha i, \beta j, \gamma k, \delta l}\right)}{m(m+2)n(n+2)} \Gamma^{(4)}_{\alpha i, \beta j, \gamma k, \delta l}.
\label{projectors}
\end{gather} 

In the course of renormalization we address here the Minimal Subtraction (MS) scheme. The renormalization constants in this scheme contain only pole contributions in $\varepsilon$ and depend only on $\varepsilon$ and coupling constants:
\begin{gather}
Z_i(g_1,g_2,\varepsilon) = 1+\sum_{k =1}^{\infty} Z_i^{(k)}(g_1, g_2)\;\varepsilon^{-k}.
\label{Zi}
\end{gather}
Renormalization constants can be found from the requirement of the finiteness of renormalized two- and four-point one-particle irreducible Green functions. Another way to calculate renormalization constants is use of Bogoliubov-Parasyuk $R'$ operation:
\begin{equation}
Z_i=1 + KR' \bar{\Gamma}_i,
\end{equation} 
where $R'$ -- incomplete Bogoliubov-Parasyuk $R$-operation, $K$ -- projector of the singular part of the 
diagram and $\bar \Gamma_i$ -- normalized Green functions of the basic theory (see e.g. \cite{Vasilev,BogShirk}) defined by the following relations:
\begin{gather}
\bar{\Gamma}_1=\frac{\partial}{\partial {m^2}}\Gamma^{(2)}\mid_{p=0},  \quad \bar{\Gamma}_2=\frac{1}{2}\left(\frac{\partial}{\partial p}\right)^2\Gamma^{(2)}\mid_{p=0}  \quad \bar{\Gamma}_3=\frac{1}{g_1 \mu^{\varepsilon}}\Gamma^{(4)}_1\mid_{p=0}, \quad 
\bar{\Gamma}_4=\frac{1}{ g_2 \mu^{\varepsilon}}\Gamma^{(4)}_2\mid_{p=0}\;.
\end{gather}

The undoubted advantage of the Bogoliubov-Parasyuk approach is that counterterms of the diagrams computed for O(1)-symmetric (scalar) model can be easily generalized to any theory with different symmetries due to the opportunity to factorize the tensor structures (see e.g. \cite{antonov2013critical,antonov2017critical,KalagovKompanietsNalimov:U(r)}). To calculate tensor factors for particular diagrams of the $O(m)\times O(n)$-symmetric model \eqref{H} one should apply projectors \eqref{projectors} to it. To implement these calculations for sufficiently large number of diagrams, we resort to the effective package of tensor algebra \textit{FORM} \cite{Vermaseren:NewFORM} and manipulating graphs package \textit{Graphine/GraphState} \cite{BatkovichKirienkoKompanietsNovikov:GraphState} while counterterm values can be taken from data obtained in the course of recent 6-loop calculations for $O(n)$-symmetric model~\cite{KP17,BCK16,KompanietsPanzer:LL2016}.
\section{RG functions, fixed points, critical exponents and marginal dimensionalities}

In this section we present expressions for RG functions, i.e. $\beta$ functions and anomalous dimensions $\gamma_{\varphi}$, $\gamma_{m^2}$, fixed-point coordinates, critical exponents, correction-to-scaling exponents $\omega_1$ and $\omega_2$ and also calculate marginal dimensionalities which determine what regime of critical behavior is realized for particular values $(m,n)$.

With the series for renormalization constants in hand, the RG functions can be calculated by means of the following relations:
\begin{gather}
\beta_i(g_1,g_2,\varepsilon) =\mu \frac{\partial g_i}{\partial\mu} \mid_{g_{01},g_{02}} = -g_i\left[\varepsilon -  g_1 \frac{\partial Z_{g_i}^{(1)}}{\partial {g_1}}-g_2 \frac{\partial Z_{g_i}^{(1)}}{\partial{g_2}}\right], \quad i = 1,2,\\
\gamma_j(g_1,g_2) = \mu \frac{\partial \log Z_j}{\partial \mu}\mid_{g_{01},g_{02}} = -  g_1 \frac{\partial Z_j^{(1)}}{\partial {g_1}}-g_2 \frac{\partial Z_j^{(1)}}{\partial{g_2}} , \quad j = \varphi, m^2,
\label{sl11}
\end{gather}
where $Z^{(1)}_i$ -- coefficients at first pole in $\varepsilon$ from \eqref{Zi}. By using these formulas we obtained the RG functions as series in terms of renormalized couplings up to six-loop order. The expansion coefficients were found analytically. Due to the cumbersomeness of six-loop expansions and the fact that they do not represent themselves any physical interest we shall limit ourselves here by presentation of the first terms. So, $\beta$ functions and anomalous dimensions in the one-loop approximation read:
\begin{gather}\label{12}
\beta_1 = -\varepsilon g_1 +\frac{(m n+8)}{3}g_1^2 -\frac{2(m-1)(n-1)}{3}g_1g_2 +\frac{(m-1)(n-1)}{3} g_2^2 +\bigo{g_i^3}, \nonumber \\
\beta_2 = -\varepsilon g_2 +\frac{(m+ n-8)}{3}g_2^2 +4g_1g_2 +\bigo{g_i^3}.
\end{gather}
\begin{gather}
\gamma_{\varphi} =\frac{(m n+2)}{18}g_1^2 -\frac{(m-1)(n-1)}{9}g_1g_2 +\frac{(m-1)(n-1)}{12} g_2^2+\bigo{g_i^3},\\
\gamma_{m^2} = -\frac{2(m n + 2)}{3}g_1 -\frac{2(m+n-mn-1)}{3}g_2 +\bigo{g_i^3}.
\label{13}
\end{gather}
The full expansions up to six-loop order for $\beta$ functions and anomalous dimensions $\gamma_{\varphi}$, $\gamma_{m^2}$ are presented in supplementary materials as \textit{Mathematica}-file (\ref{app:suppl}).

As was said above the critical behavior of the systems described by $O(m)\times O(n)$-symmetric LW action is governed by one of the four fixed points $(g_1^*,g_2^*)$ of RG equations that are zeroes of $\beta$ functions:
\begin{gather}
\beta_1(g_1^*,g_2^*,\varepsilon)=0,\qquad \beta_2(g_1^*,g_2^*,\varepsilon)=0.
\label{beta0}
\end{gather}
Solving these equations iteratively, we can find solutions as series in powers of $\varepsilon$. In one-loop approximation for arbitrary values of $n$ and $m$ the $\varepsilon$ expansions for the coordinates of chiral 
(c) and antichiral (ac) FPs are as follows: 
\begin{gather}
g_{1,\substack{c \\ ac}}^* = \varepsilon\frac{3\left(m^2 (2 n-1)+m \left(2 n^2+6 n-22\right)-n^2\pm(m+n-8)R- 22n+72)\right)}{2 \left(m^3 n+2 m^2 \left(n^2+4 n-8\right)+m \left(n^3+8 n^2-16 n-56\right)-8 \left(2 n^2+7 n-58\right)\right)}+ \bigo{\varepsilon^2},\nonumber\\
g_{2,\substack{c \\ ac}}^* =-\varepsilon\frac{3 (4+10 n-m^2 n-m (-10+4 n+n^2)\pm6 R)}{m^3 n+2 m^2 \left(n^2+4 n-8\right)+m \left(n^3+8 n^2-16 n-56\right)-8 \left(2 n^2+7 n-58\right)}+ \bigo{\varepsilon^2},
\label{15}
\end{gather}
where $R=\sqrt{m^2-2m(5 n+2)+n^2-4n+52}$. As in the case of $\beta$ functions, the fixed-point coordinates themselves are not essential.
The purpose to present here the one-loop expressions for FP coordinates is to give an idea about the restrictions which are imposed on the choice of $m$ and $n$ when calculating critical exponents of the chiral class of universality. In particular, under $m\geqslant1$ parameter $n$ has to satisfy the conditions:
\begin{gather}\label{eq:eq18}
    n\leqslant2 + 5 m - 2\sqrt{6(m^2 + m - 2)}\quad \text{or}\quad n \geqslant 2 + 5 m + 2\sqrt{6(m^2 + m - 2)}.
\end{gather}
This fact being a feature of the $\varepsilon$ expansion creates some difficulties when considering values 
of $m$ interesting from physical point of view. To solve this problem in the following sections we address some trick suggested in \cite{PELISSETTO2001605}.

Returning to the goals of this section, in order to characterize the chiral class of universality we present here the definitions of critical exponents $\alpha$, $\beta$, $\gamma$, $\eta$, $\nu$ and $\delta$. They can be expressed via anomalous dimensions $\gamma_{m^2}^*\equiv\gamma_{m^2}(g_1^*,g_2^*)$ and $\gamma_{\varphi}^*\equiv\gamma_{\varphi}(g_1^*,g_2^*)$ in the following way: 
\begin{gather}
\alpha = 2-\frac{D}{2+\gamma_{m^2}^*} , \qquad \beta  = \frac{D/2 - 1 + \gamma_{\varphi}^*}{2 + \gamma_{m^2}^*} ,\qquad \gamma  = \frac{2 - 2\gamma_\varphi^*}{2 + \gamma_{m^2}^*} , \qquad \eta = 2\gamma_\varphi^*, \nonumber \\ \qquad \nu = \frac{1}{2+\gamma_{m^2}^*}, \qquad \delta  = \frac{D+2-2\gamma_{\varphi}^*}{D-2+2\gamma_{\varphi}^*}. \label{abgd}
\end{gather}
As is well known the critical exponents are related to each other by well-known scaling relations and only two of them may be referred to as independent.

Whether the fixed point is stable or not depends on the eigenvalues $\omega_1$, $\omega_2$ of the matrix
\begin{eqnarray}\label{omega1_omega2}
\Omega=\begin{pmatrix}
\dfrac{\partial\beta_1(g_1, g_2)}{\partial{g_1} } & \dfrac{\partial\beta_1(g_1, g_2)}{\partial{g_2}}\\[1.4em] \dfrac{\partial\beta_2(g_1, g_2)}{\partial{g_1}} & \dfrac{\partial\beta_2(g_1, g_2)}{\partial{g_2}}
\end{pmatrix}
\end{eqnarray}
taken at $g_1=g_1^*$, $g_2=g_2^*$. A fixed point is stable only if both eigenvalues are positive.

As was already mentioned the numerical values of the marginal dimensionalities $n^+$, $n^-$ and $n^H$ are of the highest importance for the problem. Let us formulate the conditions to obtain these quantities. Everywhere below $m$ is considered as a fixed parameter. First, to derive series for $n^{\pm}$ it is enough to impose the following conditions in all known orders in $\varepsilon$:
\begin{gather}
\beta_1(g_{1,\pm}^*(\varepsilon),g_{2,\pm}^*(\varepsilon),n^{\pm}(m,4-\varepsilon),\varepsilon)=0, \quad \beta_2(g_{1,\pm}^*(\varepsilon),g_{2,\pm}^*(\varepsilon),n^{\pm}(m,4-\varepsilon),\varepsilon)=0,\label{21} \\
\text{det}\left|\frac{\partial(\beta_1,\beta_1)}{\partial(g_1,g_2)} \right|(g_{1,\pm}^*(\varepsilon),g_{2,\pm}^*(\varepsilon),n^{\pm}(m,4-\varepsilon),\varepsilon)=0.
\label{22}
\end{gather}
As an alternative way one may require instead of \eqref{22} the coincidence of coordinates of the chiral and antichiral FPs. In case of $n^H$ the conditions being effective from computational point of view have the following form:
\begin{gather}
\beta_1(g_{1,-}^*(\varepsilon),g_{2,-}^*(\varepsilon)=0,n^{H}(\varepsilon),\varepsilon)=0, \quad \beta_2(g_{1,-}^*(\varepsilon),g_{2,-}^*(\varepsilon)s=0,n^{H}(\varepsilon),\varepsilon)=0.\label{23}
\end{gather}
Thus it suffices to require zeroing of the second coordinate of the antichiral fixed point.
\section{Numerical results}
\subsection{Marginal dimensionalities $n^+(m,4-\varepsilon)$,
$n^-(m,4-\varepsilon)$ and $n^H(m,4-\varepsilon)$}
In this section we analyze the series for all marginal dimensionalities under the physically interesting values $m=\{2,..,6\}$. To get the proper numerical estimates from these expansions being in fact asymptotic we have to apply various resummation techniques among which are simple method of Pad\'e approximants and those based upon the Borel transformation. We pay special attention to $n^+(m=\{2,3\},4-\varepsilon)$ because these quantities are of prime physical importance. To give an idea about the numerical structure of corresponding expansions for mentioned values of $m$ we present them with the coefficients in decimals in Table \ref{tab:expansions_for_nc}.
\begin{table}[h!]
  \begin{center}
    \caption{The $\varepsilon$ expansions of marginal dimentionalities $n^+(m,4-\varepsilon)$, $n^-(m,4-\varepsilon)$, and $n^H(m,4-\varepsilon)$ for physically interesting values of $m$: $\{2,\dots,6\}$.} 
    \label{tab:expansions_for_nc}
    \setlength{\tabcolsep}{3.1pt}
    \begin{tabular}{cl}
      \hline
      \hline      
        $n^+(2,4-\varepsilon)=$&$21.798-23.431\varepsilon+7.0882\varepsilon^{2}-0.0321\varepsilon^{3}+4.2650\varepsilon^{4}-8.4436\varepsilon^{5}+\bigo{\varepsilon^{6}}$\\
        $n^-(2,4-\varepsilon)=$&$2.2020-0.5691\varepsilon+0.9892\varepsilon^{2}-2.2786\varepsilon^{3}+6.5406\varepsilon^{4}-21.696\varepsilon^{5}+\bigo{\varepsilon^{6}}$\\
        $n^H(2,4-\varepsilon)=$&$2.0000-1.0000\varepsilon+1.2942\varepsilon^{2}-2.9372\varepsilon^{3}+8.4135\varepsilon^{4}-28.311\varepsilon^{5}+\bigo{\varepsilon^{6}}$\\
        \hline
        $n^+(3,4-\varepsilon)=$&$32.492-33.719\varepsilon+11.100\varepsilon^{2}-2.1440\varepsilon^{3}+5.2756\varepsilon^{4}-8.4830\varepsilon^{5}+\bigo{\varepsilon^{6}}$\\
        $n^-(3,4-\varepsilon)=$&$1.5081-0.2816\varepsilon+0.5827\varepsilon^{2}-1.4192\varepsilon^{3}+4.0193\varepsilon^{4}-13.086\varepsilon^{5}+\bigo{\varepsilon^{6}}$\\
        $n^H(3,4-\varepsilon)=$&$1.3333-0.6667\varepsilon+0.8628\varepsilon^{2}-1.9581\varepsilon^{3}+5.6090\varepsilon^{4}-18.874\varepsilon^{5}+\bigo{\varepsilon^{6}}$\\
        \hline    
        $n^+(4,4-\varepsilon)=$&$42.785-43.939\varepsilon+14.379\varepsilon^{2}-2.9445\varepsilon^{3}+5.8589\varepsilon^{4}-8.3568\varepsilon^{5}+\bigo{\varepsilon^{6}}$\\
        $n^-(4,4-\varepsilon)=$&$1.2154-0.0607\varepsilon+0.2086\varepsilon^{2}-0.7542\varepsilon^{3}+2.2585\varepsilon^{4}-6.6644\varepsilon^{5}+\bigo{\varepsilon^{6}}$\\
        $n^H(4,4-\varepsilon)=$&$1.0000-0.5000\varepsilon+0.6471\varepsilon^{2}-1.4686\varepsilon^{3}+4.2068\varepsilon^{4}-14.156\varepsilon^{5}+\bigo{\varepsilon^{6}}$\\
        \hline
        $n^+(5,4-\varepsilon)=$&$52.923-54.119\varepsilon+17.412\varepsilon^{2}-3.2917\varepsilon^{3}+6.2043\varepsilon^{4}-8.2677\varepsilon^{5}+\bigo{\varepsilon^{6}}$\\
        $n^-(5,4-\varepsilon)=$&$1.0770+0.1188\varepsilon-0.3806\varepsilon^{2}+0.5993\varepsilon^{3}+1.8055\varepsilon^{4}-25.871\varepsilon^{5}+\bigo{\varepsilon^{6}}$\\
        $n^H(5,4-\varepsilon)=$&$0.8000-0.4000\varepsilon+0.5177\varepsilon^{2}-1.1749\varepsilon^{3}+3.3654\varepsilon^{4}-11.324\varepsilon^{5}+\bigo{\varepsilon^{6}}$\\
        \hline
        $n^+(6,4-\varepsilon)=$&$62.984-64.275\varepsilon+20.325\varepsilon^{2}-3.4212\varepsilon^{3}+6.4307\varepsilon^{4}-8.2252\varepsilon^{5}+\bigo{\varepsilon^{6}}$\\
        $n^-(6,4-\varepsilon)=$&$1.0161+0.2749\varepsilon-2.5171\varepsilon^{2}+24.972\varepsilon^{3}-152.01\varepsilon^{4}-3148.6\varepsilon^{5}+\bigo{\varepsilon^{6}}$\\
        $n^H(6,4-\varepsilon)=$&$0.6667-0.3333\varepsilon+0.4314\varepsilon^{2}-0.9791\varepsilon^{3}+2.8045\varepsilon^{4}-9.4370\varepsilon^{5}+\bigo{\varepsilon^{6}}$\\
        \hline
        \hline      
    \end{tabular}
  \end{center}
\end{table}
In order to obtain three-dimensional estimates one has to evaluate these series for $\varepsilon=1$.
The dramatic rate of the coefficient's growth makes the direct summation (just equate $\varepsilon$ to unity) absolutely fruitless. The strategy of treatment of these divergent series, i. e. the technique of their resummation will be demonstrated in next sections for specific values of $m$.

\subsubsection{$m=2$: systems with noncollinear ordering}
For $m=2,3$ the coefficients of relevant expansions up to forth order in $\varepsilon$ coincide with those 
found earlier in other works \cite{PELISSETTO2001605,ANTONENKO1995161,CALABRESE2004568}.
Because of the cumbersomeness of analytic expressions for series coefficients, they were put into the Supplementary Materials (\ref{app:suppl}). As seen from (24), the series for "lower" marginal dimensionalities $n^-(2,4-\varepsilon)$ and $n^H(2,4-\varepsilon)$, being divergent, possess regular structure that may be considered as favorable for their resummation. This allowed us to obtain numerical estimates $n^-(2,3)=1.970(3)$ and $n^H(2,3)=1.462(13)$ stable with respect to addressing different resummation techniques. Regarding $n^-(2,3)$ it is necessary to make some comment. Despite the fact that new estimate is greater than previous one $1.968(1)$ \cite{CALABRESE2004568} it can not still be referred to as accurate enough because of the fact that in case $n=m=2$ there is a mapping onto the tetragonal model \cite{PhysRevB.49.15901} where nontrivial FPs have to exist. Highly likely, this shortcut reflects bad convergence of the $\varepsilon$ expansion approach in three dimensions.

Let us consider further the series for $n^+$, which possess the worst structure among others, but corresponding numerical estimates determine the type of phase transition in real systems ($n=2,3$) and therefore play a crucial role. First, we use for resummation simple method based upon Pad\'e approximants. Standard step within this approach is a construction of so-called \textit{Pad\'e triangle} which is presented in Table \ref{pade_nc_+_m2_6loop}.  
\begin{table}[h!]
\centering
\caption{Pad\'e estimates of marginal dimensionality $n^+(2,3)$ obtained from the six-loop $\varepsilon$ expansion for chiral model. Empty boxes correspond to the approximants spoiled by dangerous (positive axis) poles.}
\label{pade_nc_+_m2_6loop}
\renewcommand{\tabcolsep}{0.3cm}
\begin{tabular}{{c}|*{7}{c}}
$M \setminus L$ & 0 & 1 & 2 & 3 & 4&5 \\
\hline
0 &21.798 & -1.63299 & 5.45517 & 5.42307 & 9.6881 & 1.24454 \\
1 &10.5055 & 3.80892 & 5.42321 & 5.45493 & 6.85442 & \text{} \\
2 &7.50313 & 4.85873 & - &   -      & \text{} & \text{} \\
3 &6.31901 & 6.04597 & 5.81301 & \text{} & \text{} & \text{} \\
4 &6.10466 & - & \text{} & \text{} & \text{} & \text{} \\
5 &5.88793 & \text{} & \text{} & \text{} & \text{} & \text{} \\
\hline
\end{tabular}
\end{table} 
Our estimate given by the most reliable approximants is $5.8(8)$. So strong scattering of Pad\'e estimates is not surprising since the behavior of expansion coefficients is rather irregular. To improve the situation we can use some tricks which are usually employed for acceleration of iteration convergence. As in \cite{CALABRESE2004568}, e. g., we can find expansion for inverse quantity:
\begin{eqnarray}\label{nc_+_num_m=2_6l_inv}
\frac{1}{n^+(2,4-\varepsilon)}=0.045876+0.049313\varepsilon&+&0.038089\varepsilon^{2}+0.024975\varepsilon^{3}\nonumber\\
&+&0.0055567\varepsilon^{4}+0.0060295\varepsilon^{5}+\bigo{\varepsilon^{6}}.\
\end{eqnarray}
Constructing Pad\'e approximants for this series we arrive to estimate $6.6(1.8)$ which also suffers from huge error bar. Another way to improve estimate is to use the insight about the value of searched quantity in different spatial dimension -- $d=2$ where according to \cite{PELISSETTO2001605} $n^+(2,2)=2$. Keeping in mind this point one can reexpand the initial series for $n^+(2,4-\varepsilon)$
\begin{eqnarray}\label{nc_+_num_m=2_6l_biased_full}
n^+(2,4-\varepsilon)=2+(2-\varepsilon)(9.8990&-&6.7660\varepsilon+0.16109\varepsilon^2\nonumber\\
&+&0.064494\varepsilon^3+\left.2.1648\varepsilon^{4}-3.1394\varepsilon^{5}\right)+\bigo{\varepsilon^{6}}\nonumber\\
&=&2+\left(2-\varepsilon\right)a^+(2,4-\varepsilon)+\bigo{\varepsilon^{6}}.\nonumber\\
\end{eqnarray}
Analysis of $a^+(2,4-\varepsilon)$ expansion within Pad\'e approximant approach gives $n^+(2,3)=6.15(19)$ while analogous treatment of the series for inverse $a^+(2,4-\varepsilon)$ \begin{eqnarray}\label{nc_+_num_m=2_6l_biased_inv_part}
\frac{1}{a^+(2,4-\varepsilon)} = 0.10102 + 0.069048\varepsilon& + &0.045551\varepsilon^{2} + 0.029352\varepsilon^{3}\nonumber\\
&-& 0.0032205\varepsilon^{4} + 0.013963\varepsilon^{5} + \bigo{\varepsilon^{6}}.\
\end{eqnarray}
results in $5.73(44)$.

The numerical estimates may be considerably improved if we address more powerful Pad\'e--Borel--Leroy (PBL) resummation technique. This approach described in a good number of papers allows one to optimize the  resummation procedure by proper choice of the shift parameter $b$ (see, for example, eq. 32 in \cite{ADZHEMYAN2019332}) and relevant comments). Having performed such an optimization we obtained $b_{opt}=1.02$. PBL triangle for $n^+(2,3)$ under this value of $b$ is presented in Table \ref{pbl_nc_+_m2_6loop} 
and yields the estimate $5.8(8)$. 
\begin{table}[h!]
\centering
\caption{The PBL estimates of marginal dimensionality $n^+(2,3)$ for $b_{opt}=1.02$ obtained from the six-loop $\varepsilon$ expansion for $O(m)\times O(n)$-symmetric model. Empty boxes correspond to the approximants spoiled by dangerous poles.}
\label{pbl_nc_+_m2_6loop}
\renewcommand{\tabcolsep}{0.3cm}
\begin{tabular}{{c}|*{7}{c}}
$M \setminus L$ & 0 & 1 & 2 & 3 & 4&5 \\
\hline
0 & 21.798 & -1.63299 & 5.45517 & 5.42307 & 9.6881 & 1.24454 \\
1 & 11.7351 & 3.5162 & 5.42325 & 5.45487 & 6.95853 & \text{} \\
2 & 9.12688 & 4.59419 & - & - & \text{} & \text{} \\
3 & 8.0747 & 5.58879 & 5.58926 & \text{} & \text{} & \text{} \\
4 & 7.5677 & 5.58926 & \text{} & \text{} & \text{} & \text{} \\
5 & 7.28485 & \text{} & \text{} & \text{} & \text{} & \text{} \\
\hline
\end{tabular}
\end{table} 
The previous result extracted from the five-loop series is $5.47(7)$ but its accuracy was claimed by the 
authors as underestimated\cite{CALABRESE2004568}. We applied the same approach to the biased expansions \eqref{nc_+_num_m=2_6l_biased_full} and \eqref{nc_+_num_m=2_6l_biased_inv_part} as well. This leads to $n^+(2,3)=6.1(4)$ and $n^+(2,3)=6.1(1)$ respectively.

Another resummation procedure known to be rather effective is also based on Borel transformation but analytical continuation is constructed by means of some conformal mapping of Borel image. This transformation helps to isolate singularities which exist on $\varepsilon$ complex plane. It is important to note that Borel summability (the presence of at most countable number of singularities on negative real axis and factorial or weaker growth of coefficients) is not proved in general for $\varphi^4$ field theories.
\begin{figure}[h!]
\centering
\includegraphics[width=12cm]{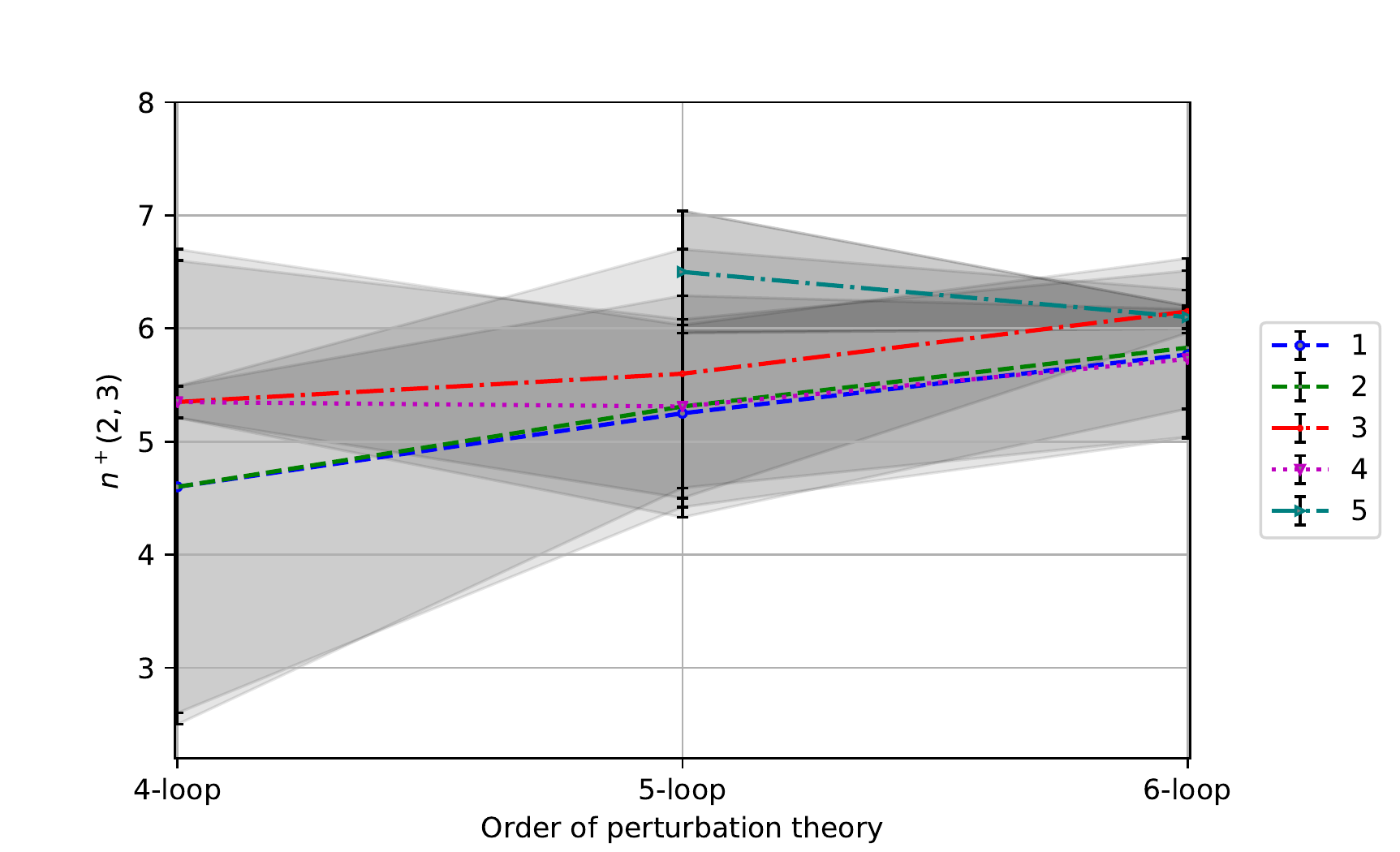}
\caption{Numerical estimates for $n^+(2,3)$ obtained by means of different resummation strategies as functions of perturbative order. 1 -- Pad\'e estimates of initial series; 2 -- PBL estimates of initial series; 3 -- Pad\'e estimates of \eqref{nc_+_num_m=2_6l_biased_full}; 4 -- Pad\'e estimates of \eqref{nc_+_num_m=2_6l_biased_inv_part}; 5 -- PBL estimates of \eqref{nc_+_num_m=2_6l_biased_inv_part}. A gray area demonstrates computational uncertainties (error bars) of numerical results obtained by means of different resummation procedures. The final estimate is $5.96(19)$.} 
\label{fig:nc2}
\end{figure}
For realization of the conform-Borel (CB) procedure it is necessary to know the asymptotic behavior of coefficients of perturbative expansions for Green functions. Such asymptotic can be obtained by means of the steepest descent method applied to action \eqref{H}\cite{PhysRevD.15.1544,Lipatov:DivergencePT}. Such an analysis for 3D $O(m)\times O(n)$-symmetric model was done in \cite{PhysRevB.68.104415}. We performed analogous analysis for $d=4-\varepsilon$ and determined corresponding singularity in $(g1,g2)$ plane, with this information it is possible to find coordinate of the singularity of the series in $\varepsilon$ (see e.g. \cite{PhysRevB.61.15136}). Coordinate of the corresponding singularity  presented in  \ref{app:asym}.  

Apart from singularity closest to the origin which defines the convergence radius for Borel image, there are other parameters which can be chosen when following the strategy suggested in \cite{KP17}. In so doing, we found the estimate $n^+(2,3)=6.0(6)$ given by CB resummation procedure. To give a general impression of all the numerical results obtained for $n^+(2,3)$ we collect in Figure \ref{fig:nc2} corresponding estimates as functions of the order of perturbation theory. The final estimate resulting from these data is $n^+(2,3)=5.96(19)$. 

\subsubsection{$m=3$: systems with noncoplanar ordering}
From the physical point of view the interest of this study is driven by the fact that under $m=3$ the model \eqref{H} describes the critical behavior in frustrated pyrochlore antiferromagnets \cite{PhysRevB.45.7295,PhysRevB.48.9881,doi:10.7566/JPSJ.87.023001,PhysRevB.94.174431,PhysRevB.91.140407,Sanders_2016,PhysRevB.92.014406}. In this case we perform all the same steps as for $m=2$.
Series coefficients for $n^{+,-}(3,4-\varepsilon)$ up to five-loop order coincide with those found earlier \cite{PELISSETTO2001605,ANTONENKO1995161,CALABRESE2004568}. 
Numerical estimates obtained for $n^H(3,3)$ and $n^-(3,3)$ by means of different resummation procedures turned out to be stable not only with respect to resummation method applied but also when moving from one perturbative order to another. All numbers obtained are presented in Table \ref{tab:comparenc}. Our final estimates for $n^-(3,3)$ and $n^H(3,3)$ are $1.408(4)$ and $0.973(11)$ respectively. Note that the estimate for $n^-$ is in a very good agreement with its five-loop counterpart $1.409(1)$ \cite{CALABRESE2004568}.

As for $n^+(3,3)$, the Pad\'e triangle with numerical estimates for initial series is presented in Table \ref{pade_nc_+_m3_6loop}. 
\begin{table}[h!]
\centering
\caption{Pad\'e estimates of marginal dimensionality $n^+(3,3)$ obtained from the six-loop $\varepsilon$ expansion for chiral model. Empty boxes correspond to the approximants spoiled by dangerous poles.}
\label{pade_nc_+_m3_6loop}
\renewcommand{\tabcolsep}{0.3cm}
\begin{tabular}{{c}|*{7}{c}}
$M \setminus L$ & 0 & 1 & 2 & 3 & 4&5 \\
\hline
0 &32.4919 & -1.22644 & 9.87379 & 7.72977 & 13.0054 & 4.52238 \\
1 &15.945 & 7.12459 & 8.07685 & 9.25425 & 9.75266 & \text{} \\
2 &11.7171 & 7.90386 & - & 9.80582 & \text{} & \text{} \\
3 &10.0051 & 9.38748 & 9.01833 & \text{} & \text{} & \text{} \\
4 &9.57092 & - & \text{} & \text{} & \text{} & \text{} \\
5 &9.21301 & \text{} & \text{} & \text{} & \text{} & \text{} \\
\hline
\end{tabular}
\end{table} 
Despite the fact that the higher-order estimates are appreciably scattered, they indicate that $n^+(3,3)$ is larger than $9$. For improving the convergence of iterations we address as earlier the PBL approach; corresponding triangle is shown in Table \ref{pbl_nc_+_m3_6loop}. Following the chosen strategy of finding the shift parameter $b$ we obtain $9.3(5)$ for $n^+(3,3)$.
\begin{table}[h!]
\centering
\caption{PBL estimates of marginal dimensionality $n^+(3,3)$ obtained from the six-loop $\varepsilon$ expansion. Empty boxes correspond to the approximants spoiled by dangerous poles. Optimal value of parameter $b$ is 15.}
\label{pbl_nc_+_m3_6loop}
\renewcommand{\tabcolsep}{0.3cm}
\begin{tabular}{{c}|*{7}{c}}
$M \setminus L$ & 0 & 1 & 2 & 3 & 4&5 \\
\hline
0 & 32.4919 & -1.22644 & 9.87379 & 7.72977 & 13.0054 & 4.52238 \\
1 &16.1995 & 7.0347 & 8.09022 & 9.23769 & 9.79091 & \text{} \\
2 &12.03 & 7.87445 & 6.44409 & 9.86726 & \text{} & \text{} \\
3 &10.3401 & 9.27327 & 8.98285 & \text{} & \text{} & \text{} \\
4 &9.74849 & 8.65446 & \text{} & \text{} & \text{} & \text{} \\
5 &9.40825 & \text{} & \text{} & \text{} & \text{} & \text{} \\
\hline
\end{tabular}
\end{table} 

Apart from the applying different resummation procedures we use the same tricks as previously to get a series with more friendly structure. First we consider the series for inverse quantity 
\begin{eqnarray}\label{nc_+_m3_inv_6l}
\frac{1}{n^+(3,4-\varepsilon)}=0.030777+0.031939\varepsilon&+&0.02263\varepsilon^{2}+0.014604\varepsilon^{3}\nonumber\\
&+&0.004534\varepsilon^{4}+0.004059\varepsilon^{5}+\bigo{\varepsilon^{6}}.\
\end{eqnarray}
The analysis of corresponding Pad\'e approximants gives $9.2(2.3)$ while evaluation by means of PBL approach leads to $\sim10.4$.

The next step we address employs the idea suggested in \cite{CALABRESE2004568} to reexpand initial series basing on the assumption that $n^+(3,2)=2$ \cite{PELISSETTO2001605}. Thus we find
\begin{eqnarray}\label{nc_+_m3_biased_6l}
n^+(3,4-\varepsilon)=2+(2-\varepsilon)(15.246&-&9.2362\varepsilon+0.93202\varepsilon^{2}\nonumber\\
&-&0.60600\varepsilon^{3}+2.3348\varepsilon^{4}-3.0741\varepsilon^{5}+\bigo{\varepsilon^{6}}\nonumber\\
&=&2+\left(2-\varepsilon\right)a^+(3,4-\varepsilon)+\bigo{\varepsilon^{6}},\nonumber\\
\end{eqnarray}
and series for inverse biased part
\begin{eqnarray}\label{nc_+_m3_biased_inv_6l}
\frac{1}{a^+(3,4-\varepsilon)}=0.065591+0.039736\varepsilon&+&0.020063\varepsilon^{2}+0.012332\varepsilon^{3}\nonumber\\
&-&0.002221\varepsilon^{4}+0.005838\varepsilon^{5}+\bigo{\varepsilon^{6}}.\
\end{eqnarray}

Pad\'e estimates for \eqref{nc_+_m3_biased_6l} and \eqref{nc_+_m3_biased_inv_6l} are $9.3(9)$ and $9.3(2)$ respectively while PBL results are $9.1(4)$ and $9.6(1.0)$.

\begin{figure}[h!]
\centering
\includegraphics[width=12cm]{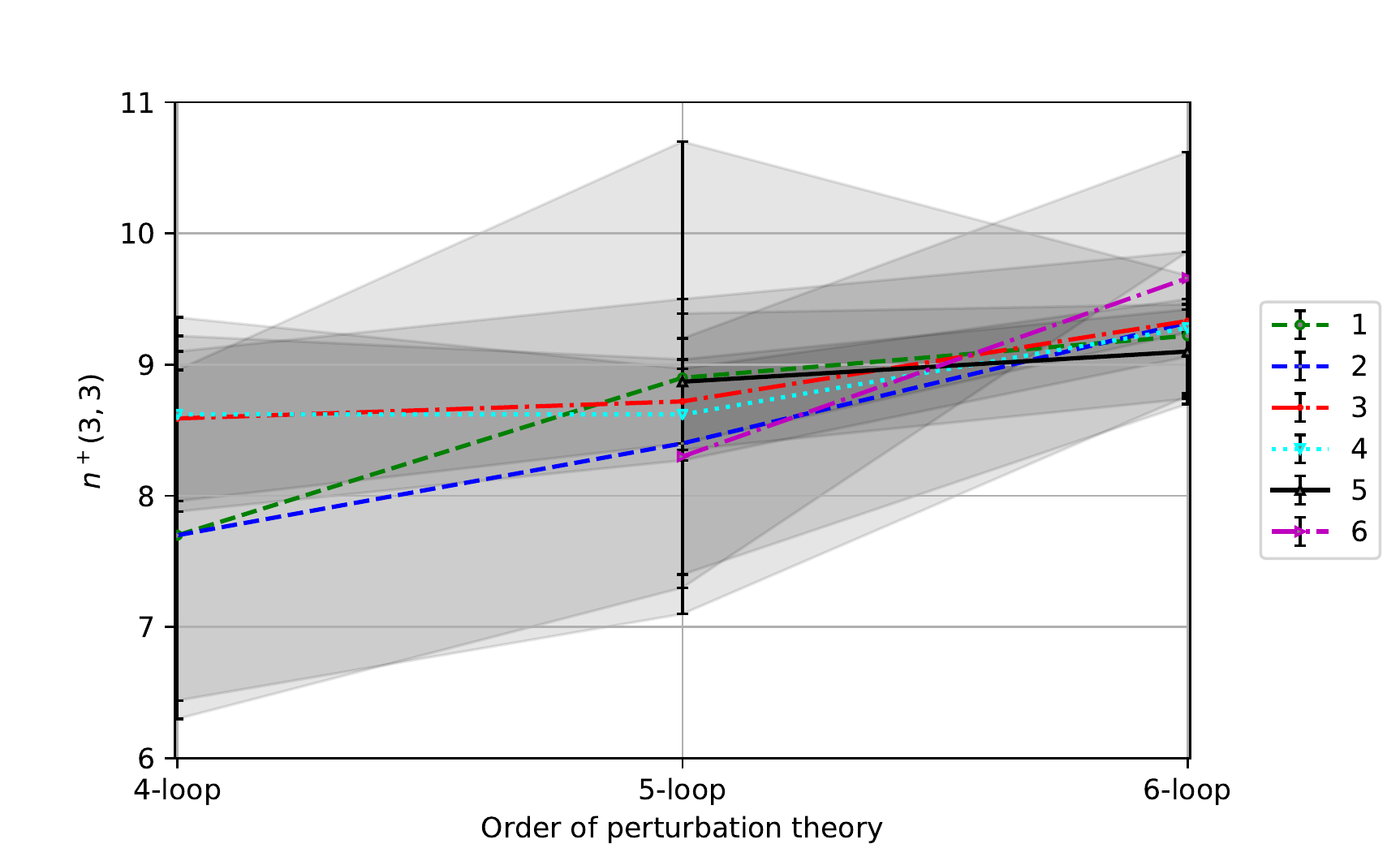}
\caption{Numerical estimates of $n^+(3,3)$ obtained by means of different resummation strategies as functions of perturbative order. 1 -- Pad\'e estimates of initial series; 2 -- PBL estimates of initial series; 3 -- Pad\'e estimates of \eqref{nc_+_m3_biased_6l}; 4 -- Pad\'e estimates of \eqref{nc_+_m3_biased_inv_6l}; 5 -- PBL estimates of \eqref{nc_+_m3_biased_6l}; 6 -- PBL estimates of \eqref{nc_+_m3_biased_inv_6l}. A gray area demonstrates error bars of numerical results obtained by means of different resummation procedures. The final value is $9.32(19)$.} 
\label{fig:nc3}
\end{figure}

The last resummation technique we apply is CB procedure. By means of this approach we found $9.3(4)$. As in  the case $m=2$ we illustrate all the numerical estimates and their accuracy in Figure \ref{fig:nc3}. The final estimate resulting from the data obtained is $n^+(3,3)=9.32(19)$. Previous field-theoretical estimates extracted from five-loop $\varepsilon$ expansion and found within six-loop 3D RG analysis are 9.5(5) \cite{CALABRESE2004568} and 11.1(6) \cite{PhysRevB.68.104415} respectively.

\subsubsection{$n^+(m,3)$, $n^-(m,3)$ and $n^H(m,3)$ for $m=\{2,..,6\}$}
In this subsection we present, without details, numerical estimates of $n^+(m,3)$, $n^-(m,3)$, and $n^H(m,3)$ for all mentioned values of $m$ including already considered. To find these estimates we implemented some of the steps which were applied previously for $m=2,3$. The only things which will be demonstrated here apart from the numbers themselves are Pad\'e and PBL triangles for the series from Table \ref{tab:expansions_for_nc} which has the most favorable structure. As an example of a such quantity we choose $n^H(5,4-\varepsilon)$. Corresponding triangles with Pad\'e and PBL estimates are presented in Table \ref{pade_nc_heis_m5_6loop} and Table \ref{pbl_nc_heis_m5_6loop} respectively.
\begin{table}[h!]
\centering
\caption{Pad\'e estimates of marginal dimensionality $n^H(5,3)$ obtained from the six-loop $\varepsilon$ expansion for chiral model.}
\label{pade_nc_heis_m5_6loop}
\renewcommand{\tabcolsep}{0.3cm}
\begin{tabular}{{c}|*{7}{c}}
$M \setminus L$ & 0 & 1 & 2 & 3 & 4&5 \\
\hline
0& 0.8 & 0.4 & 0.917695 & -0.257167 & 3.10824 & -8.21615 \\
1& 0.533333 & 0.62565 & 0.558345 & 0.613682 & 0.513842 & \text{} \\
2& 0.725373 & 0.57861 & 0.591517 & 0.576561 & \text{} & \text{} \\
3& 0.39037 & 0.594475 & 0.58276 & \text{} & \text{} & \text{} \\
4& -1.15454 & 0.557749 & \text{} & \text{} & \text{} & \text{} \\
5& 0.0909878 & \text{} & \text{} & \text{} & \text{} & \text{} \\
\hline
\end{tabular}
\end{table}

\begin{table}[h!]
\centering
\caption{PBL estimates of marginal dimensionality $n^H(5,3)$ obtained from the six-loop $\varepsilon$ expansion. The optimal value of shift parameter $b$ is $1.85$.}
\label{pbl_nc_heis_m5_6loop}
\renewcommand{\tabcolsep}{0.3cm}
\begin{tabular}{{c}|*{7}{c}}
$M \setminus L$ & 0 & 1 & 2 & 3 & 4&5 \\
\hline
0& 0.8 & 0.4 & 0.917695 & -0.257167 & 3.10824 & -8.21615 \\
1& 0.551963 & 0.611992 & 0.574068 & 0.58459 & 0.58263 & \text{} \\
2& 0.673439 & 0.586781 & 0.582275 & 0.58297 & \text{} & \text{} \\
3& 0.51544 & 0.582844 & 0.582819 & \text{} & \text{} & \text{} \\
4& 0.118464 & 0.582819 & \text{} & \text{} & \text{} & \text{} \\
5& 0.47801 & \text{} & \text{} & \text{} & \text{} & \text{} \\
\hline
\end{tabular}
\end{table} 

It is remarkable that in case of PBL procedure the numerical estimates given by two best (highest-order and near-diagonal) approximants differ from each another only in the fourth decimals. This tells us once more about the crucial role of a series structure.

\begin{figure}[h!]
\centering
\includegraphics[width=13.2cm]{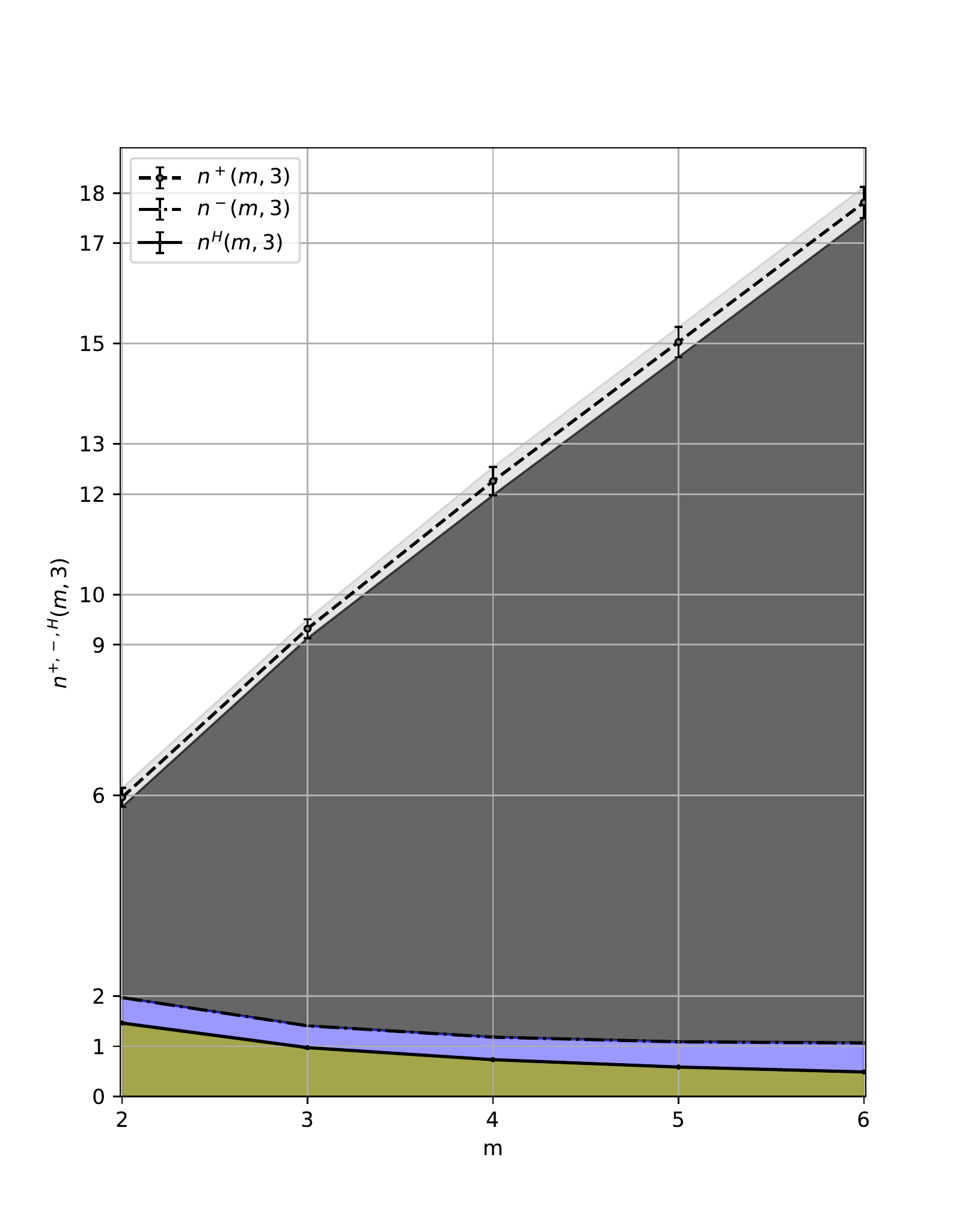}
\caption{The diagram of stability of various regimes of critical behavior including the chiral one (white area). The light gray shows the region of computational uncertainty of the lower border of the chiral critical zone. The region of fluctuation-induced first-order transitions is marked with dark gray. The olive indicates the region where the Heisenberg fixed point is stable while the area of continuous transitions into sinusoidal phase is shown as light blue.} 
\label{fig:nc+-Hall}
\end{figure}

Referring to the main goal of this section we collect the whole information concerning numerical estimates for $n^+$, $n^-$ and $n^H$ for relevant values of $m$ in Table \ref{tab:nc_all}. To get the final values we considered all estimates obtained for given quantity within the highest order of perturbation theory as the results of independent measurements. In addition, to clearly illustrate the situation we present in Figure \ref{fig:nc+-Hall} the diagram of stability of different fixed points including the chiral one in axes $(m,n)$. The empty (white) area corresponds to continuous phase transitions into chiral state while the lightest gray marks the region of computational uncertainty of its lower border.      
\begin{table}
  \begin{center}
    \caption{The numerical estimates of $n^+(m,3)$, $n^-(m,3)$, $n^H(m,3)$ for $m=\{2,\dots,6\}$ extracted from the $\varepsilon$ expansions by different resummation strategies. Notations: "$N$-loop" means that for resummation of initial series the method of Pad\'e approximants was used; "$N$-loop$_B$" indicates that Pad\'e--Borel resummation technique was applied to original expansions; "$N$-loop$_{CB}$" means that conform-Borel resummation technique was applied to initial series. B in brackets denotes that for resummation instead of initial series we take \textit{biased} one (based on a knowledge of the value at different spatial dimensionality), whereas BI means that for resummation inverse biased part of initial expansion was taken.} %
    \label{tab:nc_all}
    \begin{tabular}{lllllll}
       \hline
       & Order & $m=2$ & $m=3$& $m=4$& $m=5$& $m=6$\\
      \hline
       \multirow{18}{*}{$n^+$} & 4-loop &$4.6(2.0)$ &$7.7(1.3)$ &$10.3(1.5)$& $12.9(1.9)$& $15.(3)$ \\
                             & 4-loop$_B$ &$\underset{(b=15)}{4.6(2.1)}$ &$\underset{(b=15)}{7.7(1.4)}$ &$\underset{(b=15)}{10.3(1.6)}$      & $\underset{(b=15)}{12.8(2.1)}   $& $\underset{(b=15)}{15.3(2.8)}$ \\
                             & 4-loop(B) &$5.35(14)$ &$8.6(6)$ & &&\\
                             & 4-loop(BI) &$5.35(14) $ &$8.6(7)$ & &&\\
                             & 5-loop &$5.3(8)$ &$8.9(1.8)$& $12.(2)$& $15.(2)$ &$17.(3)$  \\
                             & 5-loop$_B$ &$\underset{(b=2.07)}{5.3(7)}$ &$\underset{(b=0.56)}{8.4(1.1)}$& $\underset{(b=0.64)}{11.2(1.3)}$& $\underset{(b=1.06)}{13.9(1.5)}$ & $\underset{(b=15)}{16.6(1.7)}$\\
                             & 5-loop(B) &$5.6(1.1)$ & $8.7(3)$& &&\\
                             & 5-loop(B)$_B$ &$- $ & $\underset{(b=0)}{8.9(5)}$& &&\\
                             & 5-loop(BI) &$5.3(1.0)$ &$8.6(4)$ & &&\\
                             & 5-loop(BI)$_B$ &$\underset{(b=1.2)}{6.5(5)}$ &$\underset{(b=0)}{8.3(0.9)}   $ & &&\\
                             & 6-loop &$5.8(7)$ &$9.2(5)$ & $12.2(5)$&$15.0(5)$&$17.6(5)$\\
                             & 6-loop$_B$ &$\underset{(b=1.02)}{5.8(8)}$   &$\underset{(b=15)}{9.3(5)}$& $\underset{(b=15)}{12.3(6)}$&$\underset{(b=2.51)}{14.9(6)}$&$\underset{(b=3.52)}{17.7(6)}$\\
                             & 6-loop(B) &$6.15(19)$                    & $9.33(9)$  & &&\\
                             & 6-loop(B)$_B$  &$\underset{(b=15)}{6.1(4)}$ &$\underset{(b=15)}{9.1(4)}   $ & &&\\
                             & 6-loop(BI)   &$5.7(4)$                    &$9.3(2) $ & &&\\
                             & 6-loop(BI)$_B$ &$\underset{(b=15)}{6.1(1)}$   &$\underset{(b=1.79)}{9.7(1.0)} $ & &&\\
                             & 6-loop$_{CB}$ &$6.0(6)$&$9.3(4)$ &$12.4(3)$ &$15.2(4)$ &$18.0(5)$\\
                             \hline
                             & Final &$5.96(19)$&$9.32(19)$ &$12.3(3)$ &$15.0(3)$ &$17.8(3)$\\
      \hline
      \multirow{7}{*}{$n^-$} & 4-loop &$1.95(9)$ &$1.40(3)$&$1.200(3)$ &$1.07(8)$&$1.05(3)$\\
                               & 4-loop$_B$ &$\underset{(b=1.08)}{1.96809(9)}$ &$\underset{(b=3.65)}{1.40897(1)}  $&$-$                            &$\underset{(b=15)}{1.07(8)}$&$\underset{(b=9.99)}{1.04513(1)}$\\
                               & 5-loop &$1.95(6)$ & $1.40(2)$   & $1.19(4)$ &$1.07(7)$&$1.055(10)$\\
                               & 5-loop$_B$ &$\underset{(b=1.06)}{1.96821(4)}$ & $\underset{(b=3.14)}{1.41002(1)} $&$\underset{(b=15)}{1.19(3)}$ &$\underset{(b=15)}{1.06(7)}$&$\underset{(b=9.99)}{1.045125(1)}$\\
                               & 6-loop &$1.969(9)$ &$1.409(4)$&$1.1801(16)$ &$1.091(16)$&$1.07(2)$ \\    
                               & 6-loop$_B$ &$\underset{(b=2.68)}{1.97091(2)}$ &$\underset{(b=3.14)}{1.410021(1)} $&$\underset{(b=15)}{1.184(14)}$ &$\underset{(b=0)}{1.087(10)}$&$1.065(13)$ \\
                               & 6-loop$_{CB}$ &$1.970(4)$ &$1.406(11)$&$1.182(9)$ &$-$&$-$ \\
                               \hline
                               & Final &$1.970(3)$&$1.408(4)$ &$1.182(6)$ &$1.089(9)$ &$1.066(12)$\\
      \hline
      \multirow{7}{*}{$n^H$} & 4-loop &$1.5(2)$ & $0.98(14)$ &$0.73(11)$ & $0.59(9)$&
                             $0.49(7)$\\
                             & 4-loop$_B$ &$\underset{(b=0)}{1.48(5)}$ & $\underset{(b=0)}{0.99(4)}$ &$\underset{(b=0)}{0.74(3)}$ & $\underset{(b=0)}{0.59(2)}$& $\underset{(b=0)}{0.493(18)}$\\ 
                             & 5-loop & $1.47(5)$ &$0.98(4)$ &$0.74(3)$&$0.59(2)$&$0.490(17)$\\
                             & 5-loop$_B$ & $\underset{(b=1.19)}{1.454(13)}$ &$\underset{(b=1.19)}{0.969(9)}$ &$\underset{(b=1.19)}{0.727(7)}$&$\underset{(b=1.19)}{0.582(5)}$&$\underset{(b=1.19)}{0.485(4)}$\\
                             & 6-loop &$1.47(4)$ &$0.98(3)$ & $0.737(19)$ &$0.590(15)$&$0.491(13)$\\
                             & 6-loop$_B$ &$\underset{(b=1.85)}{1.4575(17)}$ &$\underset{(b=1.85)}{0.9717(11)}$ & $\underset{(b=1.85)}{0.7287(8)}$ &$\underset{(b=1.85)}{0.5830(7)}$&$\underset{(b=1.85)}{0.4858(6)}$\\    
                             & 6-loop$_{CB}$ &$1.453(9)$ &$0.97(2)$ & $ - $ &$- $&$ -$\\
                             \hline
                             & Final &$1.462(13)$&$0.973(11)$ &$0.733(10)$ &$0.587(8)$ &$0.488(7)$\\
    \hline
    \end{tabular}
  \end{center}
\end{table}

\subsection{Critical exponents}
\subsubsection{$\eta$, $\gamma$, and $\nu$: $m=2$}
In this section we calculate the critical exponents of chiral universality class for $m=2$ and some relevant $n$. As is well known only two of the exponents are independent, others can be found by means of scaling relations. Here we consider only the most popular exponents -- $\eta$, $\gamma$, and $\nu$. In previous section we noted that to get 3D ($\varepsilon=1)$ numerical estimates of critical exponents in case of $n\lesssim 21.798$ we face the problem of complexity of fixed-point coordinates due to \eqref{eq:eq18} and consequently complexity of the exponents. To overcome this problem some trick was suggested \cite{PELISSETTO2001605}; its main idea is as follows. For $n$ in range $(n^+(2,3),12+4\sqrt{6})$ we reexpand the series for fixed-point coordinates and critical exponents after substituting $n=n^+(2,4-\varepsilon)+\Delta n$, where $\Delta n$ is fixed. Values of $\Delta n$ are determined by final six-loop estimate of $n^+(2,3)$ and the value of $n$ we want to consider. As was found in previous section $n^+(2,3)=5.96(19)$. Thus the interesting values of $n$ for our consideration are $n=6,7,\dots$.

As before, all the expansions of interest, because of their cumbersomeness, are presented in \textit{Mathematica}-files in Supplementary materials. Here we limit ourselves by considering the expansions for $n=n^+(2,4-\varepsilon)$ only, i.e. put $\Delta n=0$\footnote{The use of the reverse quantity instead of $\nu$ is traditional.}:
\begin{eqnarray}\label{nurev_6l_m=2_n=n+}
\nu^{-1}_{n=n^+}=2-0.5\varepsilon+0.028990\varepsilon^{2}+0.076678\varepsilon^{3}&-&0.047849\varepsilon^{4}\\ \nonumber
&+&0.076175\varepsilon^{5}-0.13380\varepsilon^{6}+\bigo{\varepsilon^{7}}.
\end{eqnarray}

\begin{eqnarray}\label{eta_6l_m=2_n=n+}
\eta_{n=n^+}=0.020833\varepsilon^{2}+0.017361\varepsilon^{3}+0.0060702\varepsilon^{4}&+&0.0031867\varepsilon^{5}\\ \nonumber &-&0.0018573\varepsilon^{6}+\bigo{\varepsilon^{7}}.
\end{eqnarray}
\begin{eqnarray}\label{gamma_n+_m2}
\gamma_{n=n^+}=1+0.25\varepsilon+0.037588\varepsilon^{2}-0.041246\varepsilon^{3}&+&0.000448\varepsilon^{4}\\ \nonumber
&-&0.034431\varepsilon^{5}+0.052175\varepsilon^{6}+\bigo{\varepsilon^{7}}.\
\end{eqnarray}
All the coefficients for $\nu^{-1}$ and $\eta$ up to five-loop contributions are in complete agreement with the results of earlier calculations \cite{CALABRESE2004568}. Making use of Pad\'e, PBL and CB resummation procedures we obtain numerical estimates for critical exponents. The numbers thus obtained are presented in Table \ref{tab:critical_exp_all}. \begin{table}[h!]
  \begin{center}
    \caption{The numerical estimates of critical exponents for $m=2$ and $n=\{n^+(2,3),6,7,8,16,32\}$ obtained by means of different resummation strategies applied to six-loop $\varepsilon$ expansions. "--" indicates that convergence of corresponding resummation procedure failed. It is remarkable that for $\eta$ our algorithm did not find any reliable approximant in case of PBL resummation procedure. "?" tells about inconsistency of numerical estimates taking into account error bars. RP - resummation procedure.} %
    \label{tab:critical_exp_all}
    \setlength{\tabcolsep}{3.1pt}
    \begin{tabular}{cllllllll}
       \hline
       & RP & $n^+(2,3)$ & $n=6$ & $n=7$& $n=8$& $n=16$&$n=32$\\
      \hline
        \multirow{3}{*}{$\nu$} & Pad\'e & $0.63(4)$ &$0.65(4)$ &$0.713(11)$ & $0.748(11)$ & $0.8710(18)$?& $0.953(3)$ \\
                               & PBL    & $0.63(4)$ &$0.65(3)$ &$0.713(12)$ & $0.745(8)$& $0.8241(17)$?&   -- \\
                               & CB     & $0.63(3)$ &$0.65(2)$ &$0.71(2)$ & $0.74(3)$ & $0.86(5)$  & $0.93(3)$\\
                               \hline
                               & Final  & $0.63(2)$ &$0.65(2)$ &$0.713(8)$& $0.745(11)$ & $0.850(16)$? & $0.940(17)$ \\
                               \hline
        \multirow{2}{*}{$\eta$} & Pad\'e & $0.048(5)$ &$0.048(5)$ &$0.045(5)$& $0.042(4)$& $0.0260(11) $ &$0.013(5)$ \\
                                & CB     & $0.046(3)$ &$0.046(3)$ &$0.044(3)$& $0.0408(17)$& $0.0261(9)$ &$0.0146(10)$ \\
                               \hline
                               & Final   & $0.047(3)$ &$0.047(3)$ &$0.045(3)$& $0.042(2)$& $0.0261(7)$ &$0.014(3)$ \\
                               \hline
        \multirow{3}{*}{$\gamma$} & Pad\'e & $1.23(10)$ &$1.26(6)$ &$1.40(2)$& $1.46(2)$ & $1.72(9)$?& $1.8949(10)$ \\
                               & PBL       & $1.24(7)$ &$1.27(5)$ &$1.40(2)$& $1.461(18)$ & $1.62(11)$ & -- \\
                               & CB        & $1.25(8)$ &$1.28(4)$ &$1.34(3)$& $1.46(5)$ & $1.69(10)$& $1.84(7)$ \\
                               \hline
                               & Final     & $1.24(5)$ &$1.27(3)$ &$1.396(14)$& $1.461(17)$ & $1.70(5)$?&$1.87(4)$ \\
                               \hline
        \multirow{3}{*}{$\omega_1$} & Pad\'e & $0.84(3)$ &$0.83(3)$ &$0.81(2)$& $0.814(17)$ & $0.860(4)$&$0.922(10)$ \\
                               & PBL         & $0.84(3)$ &$0.83(2)$ &$0.816(12)$& $0.817(10)$& $0.8599(15)$&$0.92(2)$ \\
                               & CB          & $0.85(4)$ &$0.84(6)$ &$0.81(3)$& $0.81(3)$ & $0.860(7)$&$0.92(2)$ \\
                               \hline
                               & Final       & $0.84(2)$ &$0.83(2)$ &$0.812(13)$& $0.81(4)$ & $0.860(3)$&$0.921(10)$ \\
                               \hline
        \multirow{3}{*}{$\omega_2$} & Pad\'e & $0$ & $[0.073,0.171]$ &$0.34(3)$& $0.45(2)$& $0.774(7)$& $0.90(2)$ \\
                                    & PBL    & $0$ & $[0.069,0.162]$ &$0.328(4)$& $0.44(3)$& $0.768(13)$ & $0.898(9)$\\
                                    & CB     & $0$ & $[0.071,0.167]$ &$0.34(4)$& $0.45(3)$& $0.771(12)$ & $0.909(7)$ \\
                               \hline
                                    & Final  & $0$ & $[0.069,0.171]$ &$0.34(2)$& $0.447(15)$& $0.771(6)$& $0.904(8)$ \\
                               \hline
    \end{tabular}
  \end{center}
\end{table} As was mentioned in \cite{CALABRESE2004568} addressing the trick described (substituting of shifted $n^+$ value instead of $n$) we face with extra source of errors in course of estimation of critical exponents. The scattering of numerical estimates induced by variation of resummation procedure is accompanied by an inaccuracy in determination of $n^+$ value. The first source is processed in the standard way -- by means of the averaging over all numbers obtained\footnote{These numbers -- obtained with the help of different resummation techniques -- are considered as measured independently from each other.}. Assuming monotonic dependence of critical exponents on $n$ we account for the second source calculating each critical exponent under $n^+(2,3)$ varying from $5.96-0.19$ to $5.96+0.19$. Let us consider particular example -- $\nu_{n=n^+}$. Analysis of Pad\'e approximants gives for $n=5.96$ and $6.15$\footnote{For $n=n^+$ and $n=6$ we do not calculate negative shift, otherwise we would obtain complexity in expansions. For $n \geq 7$ we calculate both negative and positive shifts and choose the biggest one.} $0.630(6)$ and $0.6652(9)$ respectively, while PBL estimates are $\nu_{n=5.96}=0.6322(14)$ for $b=0.8$ and $\nu_{n=6.15}=0.6655(6)$ under $b=0.55$. Inaccuracy in brackets are dictated by a resummation method, we include it into the error estimation. According to chosen strategy for $\nu_{n=n^+}$ we arrive to Pad\'e estimate $\nu=0.63(4)$ while PBL one was found to be $0.63(4)$. All the above remarks were kept in mind when calculating the error bars for other critical exponents.    

\subsubsection{Correction-to-scaling exponents $\omega_1$ and $\omega_2$: $m=2$}
Apart from the critical exponents, the correction-to-scaling exponents $\omega_1$ and $\omega_2$ are of prime importance from the physical point of view. As was already said these quantities determine the stability of the fixed point. Basing upon the definition \eqref{omega1_omega2} and the trick used above we can calculate corresponding $\varepsilon$ expansions for any value of $n\geqslant n^+(2,3)$. Taking into account the remarks made in previous section we found numerical estimates for $\omega_1$ and $\omega_2$. They are presented, along with those for critical exponents, in Table \ref{tab:critical_exp_all}. It is necessary to give some comment concerning $\omega_2(n=6)$. Following the adopted strategy we apply all the resummation procedures to expansions calculated at $n=6$ itself and at $6.19$ in order to take into account the inaccuracy of determination of $n^+(2,3)$. Since the values of $\omega_2$ in corresponding points are very close to zero, even small variation of $n$ can lead to noticeable scattering of numerical estimates. One can see this for $\omega_2(n=6)$, where we present the range within which the genuine value of $\omega_2(n=6)$ has to be.
\section{Discussion}
Having calculated the numbers of interest we can compare them with the results obtained by means of different field-theoretical approaches and also with estimates found in lower-order perturbative orders. As was already said, among all marginal dimensionalities $n^+$ is most interesting from the physical point of view. We collect all the estimates of $n^+(m,3)$ for $m=\{2,3,4,5\}$ obtained by means of various approaches in Table \ref{tab:comparenc}. In fact, this Table is a continuation of Table 1 in \cite{CALABRESE2004568} supplemented by the six-loop $\epsilon$ expansion results. Note that our estimate for $n^+(2,3)$ is less than six although corresponding error bar does not exclude the conclusion that systems with $O(3)\times O(6)$ symmetry should undergo first-order phase transitions.
\begin{table}[h!]
\begin{center}
    \caption{Numerical estimates of $n^+(m,3)$ for $m={2,\dots,5}$ obtained by different field-theoretical approaches within various orders of perturbation theory.} %
    \label{tab:comparenc}
    \begin{tabular}{llllll}
      \hline
      Method & Paper& $m=2$ & $m=3$& $m=4$& $m=5$\\
      \hline
        Local potential approximation & \cite{ZUMBACH1994771}1994  &$\sim 4.7$&&&\\
        3D RG: $\bigo{g^4}$ & \cite{PhysRevB.49.15901}1994 &$3.91(1)$&&&\\
        $\varepsilon$ expansion: $\bigo{\varepsilon^3}$& \cite{ANTONENKO1995161}1995 &$3.39$&&&\\
        ERG & \cite{PhysRevLett.84.5208}2000 &$\sim 4 $&&&\\
        ERG & \cite{doi:10.1142/S0217751X01004827}2001 &$\sim 5 $&&&\\
        $1/n$ expansion: $\bigo{1/n^3}$ & \cite{PELISSETTO2001605}2001 &$5.3$&$7.3$&$9.2$&$11.1$\\
        $\varepsilon$ expansion: $\bigo{\varepsilon^3}$ & \cite{PELISSETTO2001605}2001 &$5.3(2)$&$9.1(9)$&$12.1(1)$&\\
         $1/n$ expansion: $\bigo{1/n^2}$ & \cite{GRACEY2002433}2002 &$\sim 3.24$&&&\\
        3D RG: $\bigo{g^7}$  & \cite{PhysRevB.68.094415}2003&$6.4(4)$& $11.1(6)$& $14.7(8)$& $18(1)$\\
        pseudo-$\varepsilon$ expansion: $\bigo{\tau^6}$& \cite{CALABRESE2004568}2004 &$6.22(12)$& $9.9(3)$ &$13.2(6)$ &$16.3(1.3)$ \\
        $\varepsilon$ expansion: $\bigo{\varepsilon^5}$& \cite{CALABRESE2004568}2004 &$6.1(6)$ &$9.5(5)$ &$12.7(7)$ &$15.7(1.0)$\\
        $\varepsilon$ expansion: $\bigo{\varepsilon^6}$& This work &$5.96(19)$ &$9.32(19)$ &$12.3(3)$ &$15.0(3)$\\
    \hline
    \end{tabular}
  \end{center}
\end{table}

Analogously, we aggregate the known estimates for the critical exponents in Table \ref{tab:critical_exp_all_compare}. The table does not contain the results for $n=6$ because in previous works this case was not considered as actual because corresponding estimates of $n^+(2,3)$ were found to exceed 6. However for this value of $n$, besides estimates obtained in current paper, there are numerical results obtained within ERG and MC approaches. For critical exponent $\nu$ the methods of ERG and MC gave $0.707$ and $0.700(11)$ respectively while our estimate is $0.648(19)$. For the exponent $\gamma$ these techniques yielded $1.377$ and $1.38(4)$ whereas our estimate is $1.27(3)$.

For $n\geqslant n^+(2,3)$ our results are close to their counterparts obtained within the lower order in $\varepsilon$. In some cases a noticeable discrepancy with the estimates obtained by means of 3D RG and $1/n$ expansion is preserved. It is seen that the difference between the $1/n$ expansion estimates and others decreases with growing $n$ what looks quite natural since the area of applicability of $1/n$ expansion is large values of $n$. Concerning the differences between 3D RG and $\varepsilon$ expansion estimates it is worthy to recollect  that in physical dimensions (3D) both expansion parameters -- renormalized coupling and $\varepsilon$ -- are  not small preventing relevant iterations from perfect convergence. 
\begin{table}[h!]
  \begin{center}
    \caption{Numerical estimates of critical exponents for $m=2$ and $n=\{n^+(2,3),6,7,8,16,32\}$ obtained by means of different field-theoretical approaches. This Table may be considered a continuation of the Table 5 of paper \cite{CALABRESE2004568}.} %
    \label{tab:critical_exp_all_compare}
    \setlength{\tabcolsep}{1.5pt}
    \begin{tabular}{llllllll}
       \hline
       & Method &Paper& $n^+(2,3)$ & $n=7$& $n=8$& $n=16$&$n=32$\\
      \hline
        \multirow{4}{*}{$\nu$} & 3D RG: $\bigo{g^7}$  &\cite{PhysRevB.68.094415}2003                & --          & $0.68(2)$   & $0.71(1)$   & $0.863(4)$   & $0.936(1) $ \\
                               & $1/n$ expansion: $\bigo{1/n^2}$&\cite{PELISSETTO2001605}2001               & --          & $0.697$     & $0.743$     & $0.885$      & $0.946    $ \\
                               & $\varepsilon$ expansion: $\bigo{\varepsilon^5}$&\cite{CALABRESE2004568}2004& $0.635(4)$  & $0.71(4)$   & $0.75(4)$   & $0.89(4)$    & $0.94(2)  $ \\
                               & $\varepsilon$ expansion: $\bigo{\varepsilon^6}$&This work                       & $0.63(2)$ & $0.713(8)$& $0.745(11)$ & $0.850(16)$  & $0.940(17)$ \\
                               \hline
        \multirow{4}{*}{$\gamma$}& 3D RG: $\bigo{g^7}$  &\cite{PhysRevB.68.094415}2003         & --          & $1.31(5)$   & $1.40(2)$   & $1.70(1)$    & $1.860(5) $ \\
                              & $1/n$ expansion: $\bigo{1/n^2}$&\cite{PELISSETTO2001605}2001           & --          & $1.36$      & $1.45$      & $1.75$       & $1.88     $ \\
                              & $\varepsilon$ expansion: $\bigo{\varepsilon^5}$&\cite{CALABRESE2004568}2004& $1.25(2)$   & $1.39(6)  $ & $1.45(6)$   & $1.75(4)$    & $1.87(4)  $ \\
                              & $\varepsilon$ expansion: $\bigo{\varepsilon^6}$&This work                     & $1.241(48)$ & $1.396(14)$ & $1.461(17)$ & $1.70(5)$  & $1.87(4)$ \\
                              \hline
        \multirow{4}{*}{$\omega_1$} & 3D RG: $\bigo{g^7}$  &\cite{PhysRevB.68.094415}2003          & --          & $0.83(2)$   & $0.83(2)$   & $0.876(4)$    & $0.933(2) $ \\
                              & $1/n$ expansion: $\bigo{1/n^2}$&\cite{PELISSETTO2001605}2001             & --          & $0.768  $   & $0.797$      & $0.899$      & $0.949    $ \\
                              &$\varepsilon$ expansion: $\bigo{\varepsilon^5}$&\cite{CALABRESE2004568}2004& $0.86(3)  $ & $0.84(3)$   & $0.84(3)$   & $0.86(1)$    & $0.91(2)  $ \\
                              & $\varepsilon$ expansion: $\bigo{\varepsilon^6}$&This work                       & $0.840(20)$ & $0.812(13)$ & $0.81(4)$ & $0.860(3)$ & $0.921(10)$ \\
                              \hline
        \multirow{4}{*}{$\omega_2$}& 3D RG: $\bigo{g^7}$  &\cite{PhysRevB.68.094415}2003           & --  & $0.23(5)  $   & $0.36(4)$   & $0.714(9)$    & $0.868(2) $ \\
                              & $1/n$ expansion: $\bigo{1/n^2}$&\cite{PELISSETTO2001605}2001             & --  & $0.537    $   & $0.594$     & $0.797$       & $0.899    $ \\
                              & $\varepsilon$ expansion: $\bigo{\varepsilon^5}$&\cite{CALABRESE2004568}2004& $0$ & $0.33(10) $   & $0.45(8)$   & $0.77(2)$     & $0.90(1)   $ \\
                              & $\varepsilon$ expansion: $\bigo{\varepsilon^6}$&This work                      & $0$ & $0.34(2)$   & $0.447(15)$ & $0.771(6)$  & $0.904(8)$ \\
                              \hline
    \end{tabular}
  \end{center}
\end{table}

\section{Conclusion}
To sum up, we have calculated six-loop RG expansions for $O(n)\times O(m)$-symmetric (chiral) model in $4-\varepsilon$ dimensions. The series of record length together with various resummation procedures have allowed to obtain advanced numerical results for physically interesting quantities including marginal dimensionalities $n^+$, $n^-$, $n^H$ that separate different regimes of critical behavior and determine the order of phase transitions in concrete systems. The estimate for $n^+(2,3)$ found above is $5.96(19)$ indicating that transitions into chiral phase in real magnets with noncollinear ordering ($n = 2, 3$) should be first-order while the chiral class of universality is appropriate to models with $n=6$ and bigger. This conclusion is in agreement with the results of MC simulations \cite{loison2000} and ERG analysis \cite{doi:10.1142/S0217751X01004827}. For systems with noncoplanar ordering we have obtained $n^+(3,3) = 9.32(19)$ enabling one to conclude that chiral phase transitions in such materials should be first-order too. The inaccuracies of the estimates for $n^+(m,3)$ have turned out to be not small even in the highest-order approximation available what is caused mainly by rather unfavorable structure of corresponding series. Nevertheless, these estimates definitely exclude the possibility of continuous transitions into chiral phases in helical magnets and frustrated antiferromagnets. The six-loop $\varepsilon$ expansions for chiral critical exponents under $n \ge 6$ have been also derived and corresponding numerical estimates have been found.
\section*{Acknowledgment}
This work has been supported by Foundation for the Advancement of Theoretical Physics "BASIS" (grant 18-1-2-43-1). A.K. is especially grateful to Professor Pasquale Calabrese for his hospitality during the stay at SISSA where the part of the work was done.

\appendix
\section{Asymptotics of $\varepsilon$ expansions for chiral model}\label{app:asym}
In order to resum asymptotic series, one needs to know large order behavior of their coefficients. Having obtained the series for $g_1$ and $g_2$ in terms of $\varepsilon$ and following suggested in \cite{PhysRevB.61.15136} idea about determining the closest to the origin singularity $\varepsilon_b$ of $\varepsilon$ expansions we found for the chiral model: 
\begin{align}
&\frac{1}{\varepsilon_b}=-g_{1,1}^*, \quad \text{for} \quad 0<\frac{g_{2,1}^*}{g_{1,1}^*}<\frac{2m}{(m-1)}, \\
&\frac{1}{\varepsilon_b}=-\left[g_{1,1}^*-(1-\frac{1}{m})g_{2,1}^*\right], \quad \text{for} \quad \frac{g_{2,1}^*}{g_{1,1}^*}>\frac{2m}{(m-1)} \quad \text{or} \quad \frac{g_{2,1}^*}{g_{1,1}^*}<0,
\end{align}
where $g_{1,1}^*$ and $g_{1,1}^*$ -- first-order coefficients of $\varepsilon$ expansions for the fixed-point coordinates. Taking into account suggested in \cite{PELISSETTO2001605} trick (substitution of  $n=n^+(4-\varepsilon,m)+\Delta$ instead of $n$), these coefficients are as follows
\begin{align}
g_{1,1}^*&=\frac{1}{D_1}\left[3\left(\Delta^2 (2 m-1)+\Delta \left(\sqrt{\Delta \left(\Delta+4 \sqrt{6} R_1\right)}-4 \sqrt{6} R_1+2 m \left(4 \sqrt{6} R_1+11 m+2\right)-26\right)\right.\right.\nonumber \\ 
&+2\left(m\left(3 \sqrt{\Delta \left(\Delta+4 \sqrt{6} R_1\right)}+4 \sqrt{6} R_1-126\right)+\sqrt{6} R_1 \sqrt{\Delta \left(\Delta+4 \sqrt{6} R_1\right)}\right. \nonumber \\
&\left.\left.\left.-3 \sqrt{\Delta \left(\Delta+4 \sqrt{6} R_1\right)}+54 m^3+\left(22 \sqrt{6} R_1+36\right) m^2-26 \sqrt{6} R_1+36\right)\right)\right],\\
g_{2,1}^*&=\frac{1}{D_2}\left[3\left(\Delta^2 m+\Delta \left(m \left(4 \sqrt{6} R_1+11 m+8\right)-10\right)-6 \sqrt{\Delta \left(\Delta+4 \sqrt{6} R_1\right)}\right.\right.\nonumber \\
&\left.\left.-20 \sqrt{6} R_1+2 m \left(8 \left(\sqrt{6} R_1-6\right)+m \left(11 \sqrt{6} R_1+27 m+33\right)\right)-24\right)\right],
\end{align}
\begin{align}
D_1&=2\left(\Delta^3 m+\Delta^2 \left(m \left(6 \sqrt{6} R_1+17 m+14\right)-16\right)+4 \Delta \left(m \left(14 \sqrt{6} R_1\right.\right.\right.\nonumber\\
&\left.\left.+m \left(17 \sqrt{6} R_1+42 m+57\right)-69\right)-2 \left(8 \sqrt{6} R_1+15\right)\right) \nonumber \\
&\left.+12 (m-1) (m+2) \left(10 \sqrt{6} R_1+m \left(20 \sqrt{6} R_1+49 m+49\right)-44\right)\right), \\
D_2&=\Delta^3 m+\Delta^2 \left(m \left(6 \sqrt{6} R_1+17 m+14\right)-16\right)+4 \Delta \left(m \left(14 \sqrt{6} R_1 \right.\right.\nonumber \\
&\left.\left.+m\left(17 \sqrt{6} R_1+42 m+57\right)-69\right)-2 \left(8 \sqrt{6} R_1+15\right)\right)\nonumber \\
&+12 (m-1) (m+2) \left(10 \sqrt{6} R_1+m \left(20 \sqrt{6} R_1+49 m+49\right)-44\right),\\
R_1&=\sqrt{m^2+m-2}.
\end{align}
They are also presented in Mathematica file (\textit{asymptotics\_chiral.m}) as "g1" and "g2".

\section{Supplementary materials}
\label{app:suppl}
In Supplementary Materials we present expansions of RG functions for arbitrary values of $n$ and $m$. They ($\beta_1(g_1,g_2)$, $\beta_2(g_1,g_2)$, $\gamma_\phi(g_1,g_2)$, $\gamma_{m^2}(g_1,g_2)$) are put down as Mathematica file (\textit{rg\_expansion.m}). We also provide the Mathematica file with $\varepsilon$ expansions of all marginal dimensionalities under $m=\{2,\dots, 6\}$ (\textit{marg\_dim\_exp.m}). For all the couples $(m=2,n)$ considered in the paper we present the file with $\varepsilon$ expansions of critical exponents $\nu$, $\eta$, $\gamma$ and correction-to-scaling exponents $\omega_1$, $\omega_2$ describing the chiral class of universality (\textit{chiral\_crit\_exp.m}).

\newpage
\bibliographystyle{elsarticle-num}
\bibliography{chiral}
\end{document}